\documentclass[11pt]{article}
\usepackage{comment}
\usepackage[hidelinks]{hyperref}
\usepackage{enumitem}
\usepackage{amsmath,amssymb,epsf,cite,graphicx,subfigure}
\usepackage{amsmath,braket,tikzsymbols}
\usepackage[bbgreekl]{mathbbol}
\usepackage{comment}
\usepackage{soul}
\numberwithin{equation}{section}

\definecolor{airforceblue}{rgb}{0.36, 0.54, 0.66}
\newcommand{\beq}{\begin{equation}}
\newcommand{\eeq}{\end{equation}}
\newcommand{\U}{\widehat{\mathcal{U}}}
\newcommand{\nb}{\text{HH}}

\begin{document}
\baselineskip=15.5pt
\pagestyle{plain}
\setcounter{page}{1}

\begin{center}
{\LARGE \bf Emergent unitarity in de Sitter from matrix integrals}
\vskip 1cm

\textbf{Jordan Cotler$^{1,a}$ and Kristan Jensen$^{2,b}$}

\vspace{0.5cm}

{\it ${}^1$ Stanford Institute for Theoretical Physics, Stanford University, \\ Stanford, CA 94305, USA \\}

\vspace{0.3cm}

{\it ${}^2$ Department of Physics \& Astronomy, San Francisco State University, \\ San Francisco, CA 94132, USA\\}

\vspace{0.1cm}

{\tt  ${}^a$jcotler@stanford.edu, ${}^b$kristanj@sfsu.edu\\}

\medskip

\end{center}

\vskip1cm

\begin{center}
{\bf Abstract}
\end{center}
\hspace{.3cm} We study Jackiw-Teitelboim gravity with positive cosmological constant as a model for de Sitter quantum gravity. We focus on the quantum mechanics of the model at past and future infinity.  There is a Hilbert space of asymptotic states and an infinite-time evolution operator between the far past and far future. This evolution is not unitary, although we find that it acts unitarily on a subspace up to non-perturbative corrections.  These corrections come from processes which involve changes in the spatial topology, including the nucleation of baby universes.  There is significant evidence that this 1+1 dimensional model is dual to a 0+0 dimensional matrix integral in the double-scaled limit. So the bulk quantum mechanics, including the Hilbert space and approximately unitary evolution, emerge from a classical integral. We find that this emergence is a robust consequence of the level repulsion of eigenvalues along with the double scaling limit, and so is rather universal in random matrix theory. 

\newpage

\tableofcontents

\section{Introduction}

There is a wealth of evidence that quantum gravity in anti-de Sitter (AdS) spacetime is holographic, meaning that consistent theories of quantum gravity in AdS are dual to conformal field theories (CFTs), which in a sense live on the conformal boundary of AdS spacetime. This is the AdS/CFT correspondence~\cite{Maldacena:1997re}. Our understanding of these holographic dualities largely comes from string theory examples of AdS/CFT, the best studied example being the duality between maximally supersymmetric $SU(N)$ gauge theory in four dimensions and type IIB string theory on AdS$_5\times \mathbb{S}^5$. In holographic dualities, we may understand the radial direction of AdS and the fluctuations of the gravitational field as emerging from the CFT. Indeed, the only \text{known }non-perturbative definition for AdS quantum gravity is through a dual CFT when it exists.

Much less is known about quantum gravity in de Sitter (dS) space, conspicuously relevant to our existence. While there is reason to believe that consistent theories of quantum gravity are always holographic, even in de Sitter, we have much less evidence for holographic duality in dS than in AdS. The basic problem is that there are very few examples of stable dS quantum gravity to work with. The KKLT construction~\cite{Kachru:2003aw} (see~\cite{Kachru:2018aqn} for some recent comments) leads to metastable de Sitter vacua, for which it is not yet clear how to formulate a holographic correspondence. There is a positive cosmological constant version of Vasiliev theory in four dimensions~\cite{Vasiliev:1990en,Vasiliev:1999ba}, but that model is far from traditional Einstein gravity. In the absence of tractable examples there are rather basic questions about de Sitter quantum gravity that remain unanswered. For example, do we sum over complex metrics, like the Hartle-Hawking geometry? Is there a preferred quantum state of our universe, such as the no-boundary proposal? 

Questions also abound for de Sitter holography. The question of what de Sitter holography is even supposed to mean does not have an agreed upon answer, as evidenced by distinct notions of dS holography in the literature, e.g.~\cite{Hull:1998vg,Strominger:2001pn,Witten:2001kn,Alishahiha:2004md}. Perhaps the best-known one is what is usually called the ``dS/CFT correspondence''~\cite{Strominger:2001pn}, whereby quantum gravity in an inflating patch of de Sitter is dual to a non-unitary CFT on its future boundary. The best understood example of dS/CFT is the duality~\cite{Anninos:2011ui} between Vasiliev theory in dS$_4$ and the singlet sector of an $Sp(N)$ vector model. However, one expects there to be a nonzero probability to nucleate a baby universe in dS gravity, and it is not clear how this can be accommodated on the CFT side of the dS/CFT framework. Another approach is called
the ``dS/dS correspondence''~\cite{Alishahiha:2004md}. The original form of that conjecture is that de Sitter quantum gravity in $d+1$ dimensions is dual to two cutoff CFTs in $d$ dimensions, coupled together by joint sources, in addition to $d$-dimensional gravity with positive cosmological constant. This proposal has been significantly refined in the years since it was made~\cite{Dong:2010pm,Gorbenko:2018oov}, with some non-trivial tests~\cite{Gorbenko:2018oov}. Lastly, there is the viewpoint of Witten~\cite{Witten:2001kn}, which is that in de Sitter holography we should consider processes between any initial state at past infinity and any final state at future infinity, and that the output of this analysis is a Hilbert space of states. 

In this article we study one of the simplest theories of de Sitter quantum gravity in detail, namely Jackiw-Teitelboim or JT gravity~\cite{Jackiw:1984je,Teitelboim:1983ux} with positive cosmological constant. JT gravity has been the subject of enormous recent attention, beginning with~\cite{Almheiri:2014cka,Jensen:2016pah,Maldacena:2016upp,Engelsoy:2016xyb}, almost entirely with negative cosmological constant and with applications for near-extremal black holes or the Sachdev-Ye-Kitaev model~\cite{Sachdev:1992fk,kitaev} in mind. Here we build off previous work~\cite{Maldacena:2019cbz,Cotler:2019nbi} on the de Sitter version of the model (see also~\cite{Anninos:2018svg}) and are repurposing it to reliably study quantum cosmology and de Sitter holography.

JT gravity, being a model in two spacetime dimensions, has no propagating gravitons. However, it shares many features with higher-dimensional gravity, while being simple enough for us to perform concrete (and sometimes non-perturbative) computations. The model has boundary modes (an analogue of edge modes in the quantum Hall effect), moduli, and a sum over topologies. JT gravity has two coupling constants: the first is the gravitational coupling $G$, and the second a genus expansion parameter $e^{-S_0}$. The weak coupling limit is $G,e^{-S_0}\ll 1$. The gravitational coupling suppresses fluctuations of boundary modes, while fluctuations in topology are suppressed by powers of $e^{-S_0}$. For our purposes, the appealing feature of JT gravity is that we can evaluate its path integral non-perturbatively as a function of $G$ and recursively to any desired order in $e^{-S_0}$, and the resulting genus expansion is an analytic continuation of the recently computed genus expansion for JT gravity in Euclidean AdS~\cite{Saad:2019lba}. 

What about the holographic dual of JT gravity? As noted by~\cite{Saad:2019lba}, the genus expansion of JT gravity in Euclidean AdS coincides with that of a particular double-scaled matrix integral. By a similar analysis~\cite{Cotler:2019nbi}, we learn that the de Sitter genus expansion is also computed by this matrix model, and so the holographic dual of de Sitter JT gravity is a matrix integral. This duality is striking. The gravitational model is 1+1-dimensional, while on the matrix side we have a classical integral, and so are in 0+0 dimensions. In the present work, we analyze precisely how both time and space emerge from the matrix model. Furthermore, since the dual is not a CFT, nor a CFT coupled to gravity, nor a Hilbert space, this example of de Sitter holography does not appear to neatly fit into historical expectations.

As we discuss, our results appear to be most closely aligned with Witten's expectations~\cite{Witten:2001kn} for de Sitter holography. The de Sitter path integral may be understood as a transition amplitude between states prepared by the boundary conditions in the far past and far future. In this work we focus on the quantum mechanics encoded in these amplitudes. Specifically, we show that (i) non-perturbatively in $G$ and to all orders in the genus expansion, the space of asymptotic states is a Hilbert space with a non-negative norm; (ii) this norm is computed by a non-trivial path integral over large diffeomorphisms and an integral over moduli; and (iii) there is an infinite-time evolution operator from the far past to the far future, and this evolution operator is not unitary. However, (iv) to leading order in the genus expansion and \textit{exactly} as a function of $G$, the evolution operator acts as a projector, annihilating some states and as the identity on the rest. In other words the evolution operator acts unitarily on a subspace of states. We will comment on the physics of this result in a moment.  Finally, (v) unitarity is further broken at subleading order by non-perturbative effects (perturbative in powers of $e^{-S_0}$) corresponding to processes which change the topology of constant time slices. These include the creation and annihilation of baby universes.

Let us discuss the leading order result. Each component of the boundary is endowed with an edge mode, which can carry some momentum, and we can label states by this momentum. Semiclassically, initial states with positive momentum evolve into a smooth de Sitter space, while states with negative momentum contract and eventually ``crunch'' at some finite time, after which the space inflates smoothly into the future. At leading order in the genus expansion evolution simply acts as the identity on positive momentum states, and annihilates negative momentum states. 

At first glance the violation of unitarity is jarring, but with the last paragraph in mind, the leading order violation has a clear probabilistic interpretation. We can set up initial conditions that lead to a ``crunch,'' and these have zero probability to propagate to the future, while other initial conditions evolve unitarily. We also stress that JT gravity is not a manifestly unitary model, so we had no right to expect unitary evolution in the first place. In fact, there is some evidence~\cite{Saad:2019lba} that JT gravity is equivalent to a non-unitary minimal string.  From this point of view, the approximate emergent unitarity is an interesting surprise.

We also study how this emergent Hilbert space and approximately unitary evolution emerge from the dual matrix integral. Surprisingly, the mechanism in the matrix integral which gives rise to both features is rather robust and universal, namely it follows from the combination of the nearest-neighbor level repulsion of eigenvalues along with the double-scaling limit. So there is a precise sense in which double-scaled matrix integrals lead to theories of two-dimensional de Sitter quantum gravity.  We also  understand the breakdown of bulk unitarity in terms of the physics of the matrix model.  In essence, the violations of unitarity come from the low but nonzero likelihood of eigenvalues being measured outside the matrix model cut.  Along the way we show that the no-boundary state of the model is non-normalizable, and present some evidence that the Hilbert space of asymptotic states is infinite-dimensional.

The paper is organized as follows.  In Section~\ref{S:review} we review the de Sitter version of JT gravity, parameterize the space of asymptotic states, and translate the results of~\cite{Maldacena:2019cbz,Cotler:2019nbi} into a genus expansion for transition amplitudes between asymptotic states. Then in Section~\ref{S:bulkH} we obtain the inner product on the space of asymptotic states, which we find to be a Fock space. The bulk quantum gravity hands us a preferred operator, the infinite-time evolution operator from the asymptotic past to the asymptotic future, and the second is the momentum stored in the boundary modes.  We then assess the evolution operator, and show that it acts unitarily on a subspace up to non-perturbative corrections. We continue in Section~\ref{S:MM} where we show how all of these features arise from rather general considerations in the dual matrix integral.  We conclude with a discussion of our results for JT gravity and matrix integrals in Section~\ref{S:discussion}, and comment on which aspects of our analysis we expect to generalize to higher-dimensional de Sitter gravity.

\section{Nearly dS$_2$ gravity}
\label{S:review}

\subsection{Some review}

Let us briefly review Jackiw-Teitelboim (JT) gravity with positive cosmological constant (see~\cite{Maldacena:2019cbz,Cotler:2019nbi}). Its action is, up to a boundary term,
\beq
	S = S_0 \chi + \frac{1}{16\pi G} \int d^2x \sqrt{-g} \,\varphi(R-2)\,.
\eeq
The field content is a metric $g_{\mu\nu}$ and a dilaton $\varphi$. Here $\chi = \frac{1}{4\pi}\int d^2x \sqrt{-g} \, R + \frac{1}{2\pi} \int \sqrt{h} \,K$, the parameter $e^{-S_0}\ll 1$ controls the genus expansion, and we have normalized the cosmological constant to unity. The model has ``nearly'' dS$_2$ solutions, in which the spacetime is global dS$_2$ supplemented with a dilaton profile. There is a two-dimensional family of classical trajectories,
\begin{align}
\begin{split}
\label{E:globaldS}
	ds^2 &= -dt^2 + \alpha^2 \cosh^2t\, d\Psi^2\,, \qquad \Psi = \theta + \gamma f(t)\,,
	\\
	\varphi & = a\, \sinh t\,,
\end{split}
\end{align}
where $\theta\sim \theta + 2\pi$, $\gamma\sim \gamma+2\pi$, $\alpha \geq 0$, and $a$ is a constant. Also, $f(t)$ is any smooth function obeying $\lim_{t\to \pm \infty}f(t) = \pm \frac{1}{2}$. The geometry has a minimal length geodesic at time $t=0$ around the circle of length $2\pi \alpha$, and $\gamma$ indexes a ``twist'' of the future circle relative to the past circle. The analogue of the static patch has a cosmological horizon with Bekenstein-Hawking entropy $2S_0$\,. 

More generally, one integrates over configurations that respect nearly dS$_2$ boundary conditions, which characterize the boundary in terms of a signed renormalized length $\ell$.  This includes an integral over real values of the dilaton and, when the spacetime is a cylinder, over real metrics.  To define the appropriate boundary conditions, we introduce a cutoff slice close to past and future infinity, with the slice approaching infinity as $\varepsilon \to 0$.  We then fix the induced metric and dilaton on the slice to be
\beq
\label{E:BC}
	dS^2  \approx \left( \frac{\beta}{2\pi }\right)^2 \frac{d\theta^2}{\varepsilon^2} \,, \qquad
	\varphi \approx \pm\frac{1}{J\varepsilon}\,,
\eeq
where $\pm$ refers to whether one is approaching future or past infinity. In other words, the asymptotic circle has a renormalized length $\beta$, and the dilaton goes to a constant $J$ which can be positive or negative. One of the results of~\cite{Cotler:2019nbi} was that partition functions only depend on $\beta$ and $J$ through the combination $\ell = \beta J$, which may be positive or negative. Accordingly we no longer consider $\beta$ and $J$, but only the signed length $\ell$. 

We may also allow the boundary to have multiple circles, and near each we enforce the boundary conditions~\eqref{E:BC}, allowing different $\ell$'s for each boundary. The classical solutions~\eqref{E:globaldS} all have the property that $\ell_{\rm future} = \ell_{\rm past}$.

On a genus $g$ surface with $m$ future boundaries and $n$ past boundaries, $\chi$ evaluates to $-i$ times the topological Euler characteristic, $\chi = i(2g+m+n-2)$; the factor of $-i$ arises from a continuation to complex time. Integrating out the dilaton, one has a residual integral over the moduli space of constant curvature metrics. This moduli space includes a reparameterization mode on each boundary component as well as moduli. The action of the reparameterization mode is a Schwarzian action~\cite{Jensen:2016pah,Maldacena:2016upp,Engelsoy:2016xyb,Stanford:2017thb}, and its path integral is one-loop exact~\cite{Stanford:2017thb} in $G$. As a result, one may exactly evaluate the JT partition function $Z_{g,n,m}$ on a genus $g$ surface with $n$ future boundaries and $m$ past boundaries. 

JT gravity is time-reversal symmetric, and time-reversal symmetry T will play a key role in our work: parity and charge conjugation act trivially and so CPT acts as T. While the metric~\eqref{E:globaldS} of global nearly dS$_2$ space is symmetric under T, the dilaton profile is antisymmetric. It is for this reason that we introduced the $\pm$ into the dilaton part of the nearly dS$_2$ boundary condition, so T maps a past circle with some $\ell$ to a future circle with the same $\ell$.

The simplest JT partition function is that of the disk. Suppose we anchor the boundary of the disk at future infinity with some $\ell$. The result for the disk partition function is~\cite{Maldacena:2019cbz,Cotler:2019nbi}
\beq
\label{E:diskZ}
	Z_{0,1,0}(\ell)= e^{S_0}\frac{e^{\frac{\pi i}{4G \ell }}}{\sqrt{2\pi} (-2i\ell )^{3/2}}\,.
\eeq
There is also a time-reversed disk, anchored to a circle at past infinity, with $Z_{0,0,1}(\ell) = Z_{0,1,0}(\ell)^*$. Similarly one may evaluate the annulus amplitude, which includes an integral over the global dS$_2$ solutions and their moduli. The result is~\cite{Cotler:2019nbi}
\beq
\label{E:annulusZ}
	Z_{0,1,1}(\ell,\ell') = \frac{i}{2\pi} \frac{\sqrt{\ell}\sqrt{\ell'}}{\ell-\ell' + i \epsilon}\,.
\eeq
Note that this result is T-symmetric, with $Z_{0,1,1}(\ell,\ell') = Z_{0,1,1}(\ell',\ell)^*$. Beyond the disk and annulus, there is significant evidence \cite{Cotler:2019nbi} that $Z_{g,n,m}$ is an analytic continuation of the partition function $Z_{g,m+n}$ of JT gravity in Euclidean AdS on a genus $g$ surface with $m+n$ boundaries, recently obtained in~\cite{Saad:2019lba}.

\subsection{What are we computing?}

With this machinery in hand we can begin to study the observables of de Sitter JT gravity. Before doing so, it is worthwhile to go back to the more familiar setting of quantum gravity in Euclidean AdS. There, the observables are path integrals over fields in AdS, performed with fixed boundary conditions at conformal infinity. These boundary conditions include the behavior of the metric near infinity, which we usually fix to be ``asymptotically AdS.'' By the AdS/CFT correspondence the bulk path integral is mapped to a CFT path integral, and in fact the CFT gives us the only known non-perturbative definition for AdS quantum gravity. The bulk has additional, emergent spatial dimensions relative to the dual CFT.

What about quantum gravity in de Sitter space?  As noted by Witten~\cite{Witten:2001kn}, the observables are path integrals over fields in dS with fixed boundary conditions at past and future infinity.\footnote{These are not observables in the ordinary sense, since they are not accessible to a single bulk observer due to the inflationary expansion of dS. However, they ought to be computable in a holographic dual.} These are the de Sitter analogues of the usual AdS boundary conditions, which include the conformal class of the metric at infinity. However, here we encounter a departure from the usual situation in Euclidean AdS, where we have a single boundary component and a single dual CFT.\footnote{Of course, there is an old question of whether or not one should sum over Euclidean wormholes (i.e., geometries which connect multiple asymptotic regions) in Euclidean gravity. At least for JT gravity, the integral over wormhole geometries is sensible and can be mapped to a holographically dual matrix integral~\cite{Saad:2019lba}. The status of Euclidean wormholes for string theory on AdS is less clear, see e.g.~\cite{Maldacena:2004rf,ArkaniHamed:2007js}.} In de Sitter the classical trajectory has multiple boundaries, one in the past and one in the future, and in the quantum theory there does not seem to be an obstruction to summing over geometries with an arbitrary number of boundary components.

The conformal boundary of de Sitter is spacelike, and so the de Sitter path integral with these boundary conditions may be interpreted as a transition amplitude: the overlap $\braket{f|i}$ of an initial state $\ket{i}$, specified by the past boundary condition, with a similarly prepared future state $\bra{f}$. We refer to these states as ``asymptotic.'' We would like to interpret this amplitude as a matrix element of the infinite evolution operator from past to future, $\bra{f}\U\ket{i}$. However, it has not been clear how to separately form an inner product on the vector space of asymptotic states. Without such an inner product, all one seems to have are vector spaces of past and future asymptotic states, and a pairing between initial states and final states induced by bulk time evolution. Witten observed that one may use bulk CPT symmetry $\Theta$ to construct a Hermitian inner product using the evolution operator, through $(i,j) = \bra{\Theta j}\U\ket{i}$. So one may trade the evolution operator for a norm, but the upshot is that one then gets a Hilbert space of asymptotic states. 

One of the points of this paper is that one can do better, at least for JT gravity and some generalizations thereof. The asymptotic states of nearly dS$_2$ JT gravity are labeled by the number of boundary circles, and by a signed length $\ell_i$ for each boundary. Time reversal maps past states to future states. In the next Section we describe a variant of the de Sitter path integral which computes the inner product between two such states. So the space of asymptotic states is in fact a Hilbert space, and time reversal gives us a canonical isomorphism between the past and future Hilbert spaces. Moreover, the inner product is both non-degenerate and non-negative. With this inner product, we may study $\U$ and assess whether or not it is unitary.

Earlier we displayed the results for the disk and annulus partition functions of de Sitter JT gravity. Here we discuss what these results mean in terms of matrix elements of the evolution operator, which we will temporarily call $\U_0$\,. The disk has a single future circle and no past asymptotic region. The geometry smoothly caps off in the bulk. The standard interpretation is that the smooth cap prepares a state in the bulk Hilbert space at some intermediate time, the no-boundary state, and the disk computes the leading contribution to the Hartle-Hawking wavefunction of the universe in this state,
\beq
	\Psi_{\rm HH}(\ell) \simeq Z_{0,1,0}(\ell) + O(e^{-S_0})\,.
\eeq
The corrections come from higher genus geometries. This wavefunction is the overlap
\beq
	\Psi_{\rm HH}(\ell) = \braket{\ell|\nb}\,,
\eeq
where $\ket{\nb}$ is the no-boundary state, evolved to the infinite future so that it becomes a superposition of asymptotic states. At this stage $|\text{HH}\rangle$ is not normalized, and it is not clear if $\ket{\nb}$ is normalizable or not, since the $\bra{\ell}$'s have not yet been orthonormalized. We will see in the next Section that $\ket{\nb}$ is non-normalizable.  Because the unnormalized state $\ket{\nb}$ is well-approximated by the disk partition function, it is of order $e^{S_0}$.

Now consider the simplest matrix element of $\U_0$\,, between a 1-boundary state $\ket{\ell'}$ in the past and a 1-boundary state $\bra{\ell}$ in the future. It is
\beq
	\bra{\ell}\U_0\ket{\ell'} \simeq Z_{0,1,0}(\ell)Z_{0,0,1}(\ell') + O(e^{0\times S_0}) \simeq \braket{\ell|\nb}\braket{\nb|\ell'}+ Z_{0,1,1}(\ell,\ell') + O(e^{-2S_0})\,.
\eeq
The leading contribution, of $O(e^{2S_0})$, comes from a sum over geometries with the topology of two disconnected disks, which we can think of as describing the likelihood for the past universe to disappear and for the future universe to bubble. The second term is the annulus amplitude, of $O(e^{0\times S_0})$, and the third term comes from higher genus geometries.

Because the no-boundary state $\ket{\nb}$ is non-normalizable, the evolution operator $\U_0$ (defined through the sum over all geometries) therefore takes normalizable states to non-normalizable states. In order to afford a putative probabilistic interpretation, we therefore define the evolution operator $\U$ by projecting out the no-boundary state,
\beq
\label{E:projectHH}
	\U \equiv \U_0 - \ket{\nb}\bra{\nb}\,.
\eeq
The matrix elements of $\U$ between 1-boundary states are computed by the annulus amplitude plus genus corrections,
\beq
	\bra{\ell}\U\ket{\ell'} \simeq Z_{0,1,1}(\ell,\ell') + O(e^{-2S_0})\,,
\eeq
which come from higher genus spaces which connect the past circle to the future circle.

\begin{figure}
\begin{center}
\includegraphics[scale=.55]{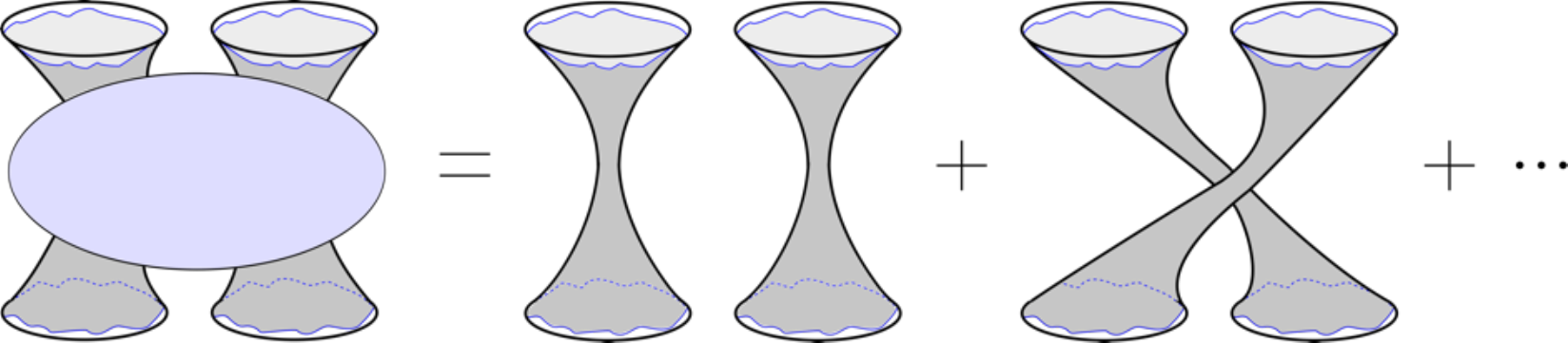}
\end{center}
\vskip.3cm
\caption{\label{fig:2to2process} The transition amplitude $\bra{\ell_1,\ell_2}\U\ket{\ell_1',\ell_2'}$ is a sum over geometries which connect two circles at past infinity with two circles at future infinity. At leading order in the genus expansion, the amplitude is a sum of two terms, analogous to Wick contractions, as given in Eq.~\eqref{E:2to2}. The dots indicate subleading terms in the genus expansion. Each boundary circle is ``wiggly,'' on account of the Schwarzian reparameterization mode living on it. }
\end{figure}

More generally, matrix elements of $\U$ are computed by the sum over geometries which connect past infinity to future infinity. This feature is crucial: it precludes the disk, although it does not forbid disconnected spacetimes. For example, consider the transition between the past state $\ket{\ell_1',\ell_2'}$ and future state $\bra{\ell_1,\ell_2}$. See Fig.~\ref{fig:2to2process}. The overlap approximately factorizes in the genus expansion as
\begin{align}
\begin{split}
\label{E:2to2}
	\bra{\ell_1,\ell_2}\U\ket{\ell_1',\ell_2'} &\simeq \bra{\ell_1}\U\ket{\ell_1'} \bra{\ell_2}\U\ket{\ell_2'}+\bra{\ell_1}\U\ket{\ell_2'}\bra{\ell_2}\U\ket{\ell_1'} + O(e^{-2S_0})
	\\
	&\simeq Z_{0,1,1}(\ell_1,\ell_1')Z_{0,1,1}(\ell_2,\ell_2') + Z_{0,1,1}(\ell_1,\ell_2')Z_{0,1,1}(\ell_2,\ell_1') + O(e^{-2S_0})\,,
\end{split}
\end{align}
and each of the two terms shown is (approximately) the partition function on two disconnected annuli. Because we sum over geometries which connect past to future, we do not include a third term at $O(e^{0\times S_0})$, namely $Z_{0,2,0}(\ell_1,\ell_2)Z_{0,0,2}(\ell_1,\ell_2')$. This term would correspond to a disconnected product of two annuli, one which glues the two future circles while smoothly capping off in the interior, and another which does the same for the two circles in the past.

We denote the most general matrix element of $\U$ as
\beq
	\bra{\ell_1,\hdots,\ell_n}\U\ket{\ell_1',\hdots,\ell_m'} \simeq Z_{n,m}(\ell_i,\ell'_j) \,,
\eeq
where $\ket{\ell_1,\hdots\ell_m}$ is the state prepared by $m$ asymptotic circles with parameters $\ell_1,\hdots \ell_m$\,, $\ket{\ell_1',\hdots,\ell_n'}$ is defined similarly, and $Z_{n,m}$ is the formal power series given to us by the sum over geometries which connect $n$ future circles to $m$ past circles in the genus expansion. For example, $Z_{1,1}(\ell,\ell') =\sum_{g=0}^{\infty} Z_{g,1,1}(\ell,\ell')$. We write $\simeq$'s in these expressions because the genus expansion is asymptotic.

\section{Asymptotic Hilbert space and infinite evolution}
\label{S:bulkH}

In the last Section we explained how the de Sitter JT path integral computes transition amplitudes from states in the asymptotic past to states in the asymptotic future.  These states were characterized by the number of connected components of the future or past boundary, and by a signed length $\ell_i$ for each component. Here, we sharpen our understanding by studying the quantum mechanics underlying these amplitudes. In particular, we would like to answer the following questions: What is the Hilbert space of asymptotic states? What is the inner product on the space? What is the bulk interpretation of operators which act on this space? Is infinite-time evolution unitary? Is the no-boundary state normalizable? We begin by re-examining the simplest amplitudes, from 1-boundary states to 1-boundary states, and recast this into the language of single-particle quantum mechanics. Then we consider the multiple-boundary case, and extend our analysis to a third-quantized framework.

\subsection{One past boundary and one future boundary}

Consider again the amplitude $\bra{\ell}\U\ket{\ell'}$ to transition from a 1-boundary state with $\ell'$ in the past to a 1-boundary state with $\ell$ in the future. Recall that $\ell$ is defined in terms of the nearly dS$_2$ boundary condition~\eqref{E:BC} as $\ell \equiv \beta J$, where $\beta$ is the renormalized length of the circle at infinity and $J$ determines the asymptotic growth of the dilaton. The renormalized length is non-negative, but the dilaton can be of either sign, and so $\ell$ can taken on any real value: it is a renormalized, signed length of the boundary. At this stage in our analysis we only have a vector space of asymptotic states $\ket{\ell}_P$ in the past and a vector space of dual states $_F\!\bra{\ell}$ in the future; the vector spaces are not yet Hilbert spaces, since we do not have an inner product. We denote these spaces as
\beq
	\mathcal{H}_P^{1\text{-bdy}} \simeq  \text{span}\{\ket{\ell}_P \}_{\ell \in \mathbb{R}}\,, \qquad \quad \mathcal{H}_F^{1\text{-bdy},*} \simeq  \text{span}\{\phantom{}_F\!\bra{\ell}\}_{\ell \in \mathbb{R}}\,.
\eeq
Before examining $\widehat{\mathcal{U}}$, we will first demonstrate that these spaces are in fact canonically isomorphic Hilbert spaces. We do so by first using the JT path integral to obtain an inner product on $\mathcal{H}_P^{1\text{-bdy}}$ and $\mathcal{H}_F^{1\text{-bdy}}$, and then using bulk time-reversal to map past to future.

In ordinary quantum mechanics we may consider use the path integral to compute matrix elements of the finite-time evolution operator, i.e. the propagator, via
\beq
	\bra{x_f}\U(t_f,t_i)\ket{x_i} = \int [dx(t)]_{x(t_i)=x_i}^{x(t_f)=x_f} e^{i S[x]}\,.
\eeq
The initial and final states are implemented by boundary conditions on $x(t)$. Since $\lim_{t_f\to t_i}\U(t_f,t_i) = \textbf{1}$, it follows that we may use the path integral to compute the inner product $\braket{x_f|x_i}$ by simply taking the limit of the path integral as $t_f\to t_i$. This computation is effectively classical, and the path integral is dominated by the straight-line trajectory, $x(t) = x_i+ \frac{t-t_i}{t_f-t_i}(x_f-x_i)$. 

We would like to adapt this procedure to JT gravity. The catch is that in a theory of gravity we cannot access the finite-time path integral. The standard path integral is the one where the initial time is in the asymptotic past, $t_i\to-\infty$, and the final time in the asymptotic future, $t_f\to +\infty$. However, this is not the only option: the gravitational path integral is sensible as long as the initial and final times are asymptotic, and so we can just as well send them both to the asymptotic past, or to the asymptotic future. We then \emph{define} $_P\!\braket{\ell|\ell'}_P$ to be the path integral in the asymptotic past with $t_f,t_i\to-\infty$, and $_F\!\braket{\ell|\ell'}_F$ to be the corresponding integral in the future.

\begin{figure}
\begin{center}
\includegraphics[scale=.5]{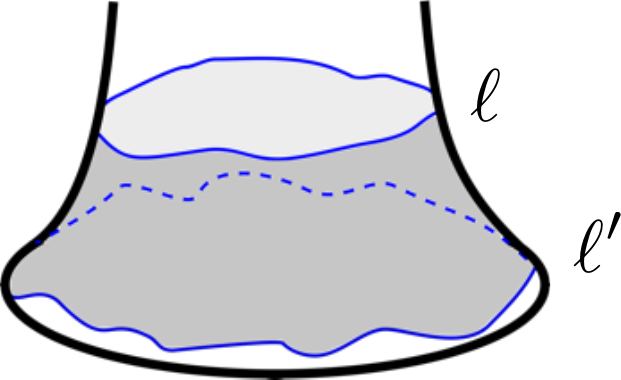}
\end{center}
\vskip.1cm
\caption{\label{fig:innerproduct} The inner product $\braket{\ell|\ell'}$ is a sum over patches of spacetime in the asymptotic past. This patch runs between a circle with renormalized signed length $\ell'$, on which we impose the past version of the nearly dS$_2$ boundary conditions~\eqref{E:BC}, and another circle with $\ell$ on which we impose the future version of the nearly dS$_2$ boundary conditions.}
\end{figure}

To leading order in the genus expansion we are integrating over annuli, patches of dS$_2$ spacetime that connect two circles in the asymptotic past. The circle with length $\ell'$ prepares the initial state, and so on that circle we impose the ``past'' version of the nearly dS$_2$ boundary conditions~\eqref{E:BC}, while on the other circle with length $\ell$ we impose the ``future'' version of the nearly dS$_2$ boundary conditions. The computation of this path integral is rather similar to that of the annulus partition function described in detail in~\cite{Cotler:2019nbi}. See Fig.~\ref{fig:innerproduct}. Let us first present the result and then explain where the pieces come from. We find
\beq
\label{E:innerproduct1}
	_P\!\braket{\ell|\ell'}_P \simeq \int_{-\infty}^{\infty} \frac{db^2}{4G}\int_0^{2\pi} d\gamma \, Z_T(\ell,b^2)Z_T(\ell',b^2)^* + O(e^{-2S_0})\,,
\eeq
where
\beq
	Z_T(\ell,b^2) = \int \frac{[df(\theta)]}{U(1)} \exp\left( \frac{i}{4G \ell}\int_0^{2\pi} d\theta\left( \{f(\theta),\theta\} - \frac{b^2}{2}f'(\theta)^2\right)\right) \,,
\eeq
is the path integral over the Schwarzian mode on the ``future'' circle, and its complex conjugate $Z_T(\ell',b^2)^*$ is the path integral over the Schwarzian mode on the ``past'' circle. The parameters $b^2$ and $\gamma$ label the moduli of the annulus, with $b^2$ related to $\alpha^2$ in Eq.~\eqref{E:globaldS} as $b^2 = - \alpha^2$. The measure over $b$ and $\gamma$ is the Weil-Petersson measure which can also be obtained from a gravitational computation~\cite{Saad:2019lba,Cotler:2019nbi}. This is nearly the same integral that computes the annulus partition function, which is given by the same integrand but over the domain $b^2 \geq 0$. To understand why we integrate over all $b^2$, consider again the metric on global dS$_2$, which in the asymptotic past reads
\begin{equation*}
	ds^2 = -dt^2 + \left( e^{-2t} -b^2 + O(e^{2t})\right) d\theta^2\,.
\end{equation*}
So as $t\to-\infty$, we get a smooth spacetime for any real value of $b^2$, and we integrate over all possible values.\footnote{For the global dS$_2$ partition function, the minimal length of the spatial circle is $2\pi \alpha$ and so naively we integrate over all $\alpha^2 \geq 0$. However the one-loop determinant has wrong-sign modes for $|\alpha|>1$, and so we define the Schwarzian path integral in that region by continuation from the domain of positive $b^2$. In that domain the future integral converges with an $i\epsilon$ prescription for $\ell$, namely $\ell \to \ell + i \epsilon$, and the past converges with $\ell' \to \ell' - i\epsilon$.  It remains to perform the moduli space integral, but when it is done over positive $\alpha^2$, the integral converges with the opposite $i\epsilon$ prescription. So we must perform a second continuation, thus obtaining the correct $i\epsilon$ prescription. Alternatively, we can define the moduli space integral by instead integrating over positive $b^2$, which converges with the right $i\epsilon$ prescription. Either way leads to the same answer for the annulus partition function~\eqref{E:annulusZ}. We are performing a similar procedure in writing the inner product~\eqref{E:innerproduct1} as an integral over $b^2$. } The Schwarzian path integral $Z_T$ is one-loop exact, with 
\beq
	Z_T(\ell,b^2) = \frac{1}{\sqrt{-4\pi i \ell}}\,e^{-\frac{i \pi b^2}{ 4G\ell}}\,,
\eeq
from which we find
\beq
	_P\!\braket{\ell|\ell'}_P \simeq \sqrt{\ell}\sqrt{ \ell'} \delta(\ell-\ell') + O(e^{-2S_0})\,.
\eeq
By renormalizing the states as $\ket{\ell}_P\to \ket{\ell}_P/\sqrt{\ell}$, we obtain the delta function normalization
\beq
	_P\!\braket{\ell|\ell'}_P \simeq  \delta(\ell- \ell') + O(e^{-2S_0})\,.
\eeq
A similar computation holds for $\mathcal{H}_F^{1\text{-bdy}}$, with the same inner product, as guaranteed by time-reversal symmetry. In fact, since time-reversal maps the past nearly dS$_2$ boundary condition with some $\ell$ to the future nearly dS$_2$ boundary condition with the same $\ell$, we see that combined with the inner product, time-reversal gives us a canonical isomorphism $\ket{\ell}_P \longleftrightarrow \ket{\ell}_F$ and so we can simply drop the $P$ and $F$ subscripts in what follows.

What are the genus corrections to the  inner product? Now we borrow a result from previous work~\cite{Maldacena:2019cbz,Cotler:2019nbi}. In the JT path integral the dilaton enforces a constant curvature constraint, $R=2$. However, there are no smooth higher genus metrics with constant positive curvature. So one has to define the JT path integral on a higher genus surface with some prescription. One can analytically continue from the Euclidean AdS version of JT gravity, as in~\cite{Maldacena:2019cbz}, or exploit a topological gauge theory formulation of JT gravity, in which one is integrating over smooth flat connections on a higher genus surface~\cite{Cotler:2019nbi}. There is significant evidence summarized in~\cite{Cotler:2019nbi} that both points of view lead to the same partition functions. Either way, there are higher genus contributions to the path integral, which may be thought of as constrained complex-time instantons. These surfaces are composed of asymptotic annuli glued to an intermediate higher genus surface in the ``middle'' of the spacetime, infinitely far from the boundary. As a result, in computing $\braket{\ell|\ell'}$ where we are integrating over surfaces near conformal infinity, the only surface which contributes is the annulus. And so we have
\beq
	\braket{\ell|\ell'} \simeq \sqrt{\ell}\sqrt{\ell'} \,\delta(\ell-\ell') + (\text{non-perturbative in }e^{-S_0})\,.
\eeq
The potential non-perturbative corrections are best called ``doubly non-perturbative,'' since the genus expansion is already a sum over non-perturbative contributions to the path integral.

In the remainder of this Subsection we are interested in de Sitter physics within the genus expansion, and so henceforth we work with the normalized 1-boundary states.

Now that we have the inner product, let us reexamine the Hartle-Hawking wavefunction of the no-boundary state $\ket{\nb}$ in terms of normalized 1-boundary states. From~\eqref{E:diskZ} it is
\beq
	\braket{\ell|\nb} \simeq \frac{Z_{0,1,0}(\ell)}{\sqrt{\ell}} +O(e^{-S_0})= e^{S_0}\,\frac{e^{\frac{i \pi}{4G \ell }}}{\sqrt{2\pi} \,\ell^2\, (-2 i )^{3/2}} \,+\, O(e^{-S_0})\,,
\end{equation}
where now the 1-boundary state $\bra{\ell}$ is normalized. The genus corrections have an absolute value which is regular at small $\ell$, and so this divergence is not ameliorated by the genus expansion. As a result, the Hartle-Hawking wavefunction, which expresses the no-boundary state as a superposition of asymptotic states, is non-normalizable to all orders in the genus expansion. It is for this reason that we define the infinite-time evolution operator $\U$ as in~\eqref{E:projectHH} by projecting out the no-boundary state.

\begin{figure}
\begin{center}
\includegraphics[width=1.3in]{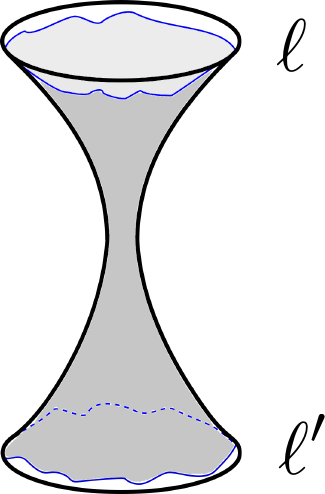}
\end{center}
\caption{\label{F:global} The basic observable of de Sitter JT gravity is the annulus partition function, $Z_{0,1,1}(\ell,\ell')$, the path integral over global nearly dS$_2$ space and metric/dilaton configurations connected to it. Properly normalized it gives the leading expression for the infinite-time evolution operator in~\eqref{E:leadingU}.}
\end{figure}

Next we may consider the transition amplitude between normalized 1-boundary states. It is
\beq
\label{eq:normalizedprop1}
	\bra{\ell}\U\ket{\ell'} \simeq \frac{Z_{0,1,1}(\ell,\ell')}{\sqrt{\ell}\sqrt{\ell'}}  + O(e^{-2S_0})= \frac{i}{2\pi} \frac{1}{\ell - \ell'+ i 
\epsilon} \, + \, O(e^{-2S_0})\,.
\eeq
This amplitude is the basic observable of de Sitter JT gravity. See Fig.~\ref{F:global}. We may regard it as the ``free'' propagator of a universe. Indeed, recalling that the classical de Sitter solution has $\ell = \ell'$, we see that the divergence of this amplitude coincides with when the universe goes on-shell.

This amplitude is emergent under a seemingly emergent ``translation'' symmetry $\ell\to \ell+c, \ell'\to \ell'+c$. So it is natural to define a ``momentum'' $p$ conjugate to $\ell$ and to work in momentum space.  We define $p$ to have opposite its usual sign.  The momentum eigenstates
\beq
	\ket{p}= \frac{1}{\sqrt{2\pi}} \int_{-\infty}^\infty d\ell \, e^{i p \ell} \ket{\ell}
\eeq
are orthonormal in the genus expansion, and in terms of them we have
\beq
	\label{eq:normalizedprop2}
	\bra{p}\U\ket{p'} \simeq \Theta(p) \, \delta(p-p') \, + \, O(e^{-2S_0})\,,
\eeq
where $\Theta(p)$ is the Heaviside step function. Clearly we may define canonically conjugate Hermitian operators $\hat{\ell}$ and $\hat{p}$, i.e.~$[\hat{\ell},\hat{p}] =- i$, on the space of 1-boundary states via $\hat{\ell}\ket{\ell} = \ell\ket{\ell}$ and $\hat{p}\ket{p} = p\ket{p}$. We will return to the physical meaning of these operators shortly. In terms of $\hat{p}$ we see that infinite-time evolution operator $\U$ acts on the space of 1-boundary states as
\beq
\label{E:leadingU}
	\U\simeq \Theta(\hat{p}) + O(e^{-2S_0})\,.
\eeq
So up to non-perturbative corrections, $\U$ acts as the identity\footnote{Time-reversal symmetry guarantees that the infinite-time evolution operator $\U$ is Hermitian. If there is a subspace $S$ on which $\U$ acts unitarily, then on that subspace $\textbf{1}_S =\U_S^{\dagger}\U_S = \U_S^2$, so that the eigenvalues of $\U_S$ are $\pm 1$.} on negative momentum states and annihilates positive momentum states. Said another way, to leading order in the genus expansion, negative momentum initial states are ``unstable'': they have zero likelihood of evolving into a de Sitter universe and propagating to the future. This gives one of our main results, that $\U$ acts unitarily up to genus corrections on the subspace of positive momentum states. Later in this Section we will see that the genus corrections lead to distributional terms in $\U$ at $p=0$, and so momentum eigenstates with $p>0$ evolve unitarily to all orders in the genus expansion.

Let us make a comment which we will return to in Section~\ref{S:MM}. The annulus partition function $Z_{0,1,1}$ can be written as
\beq
	Z_{0,1,1}(\ell,\ell') = \sqrt{\ell}\sqrt{\ell'} \left( \frac{i}{2\pi}\, \text{PV}\frac{1}{\ell-\ell'} + \frac{1}{2}\,\delta(\ell-\ell')\right)\,.
\eeq
Indeed, the inner product of 1-boundary states $\braket{\ell|\ell'}\simeq\sqrt{\ell}\sqrt{\ell'}\,\delta(\ell-\ell')$ equals the singular discontinuity in $Z_{0,1,1}$ as one takes the imaginary part of $\ell$ from negative to positive. This is a consequence of the fact that the path integral which computes the overlap is essentially the same one that computes $Z_{0,1,1}$\,, except the domain of integration is over all $b^2$ instead of $b^2\geq 0$. So in hindsight we can infer the inner product from the divergence in $Z_{0,1,1}$\,. 

Now we return to the geometric interpretation of the operators $\hat{\ell}$ and $\hat{p}$. The observable $\hat{\ell}$ measures the signed length of the boundary, but what of $\hat{p}$\,? To answer this question, it is helpful to return to the JT path integral which computes the unnormalized overlap $\braket{\ell|\ell'}$\,:
\beq
\label{E:1bdyoverlap}
	\braket{\ell|\ell'} \simeq \int_{-\infty}^{\infty} \frac{db^2}{4G}\int_0^{2\pi} d\gamma \int \frac{[df_1][df_2]}{U(1)\times U(1)} e^{\frac{i}{4G\ell}\int_0^{2\pi} d\theta \left( \{f_1(\theta),\theta\}-\frac{b^2}{2}f_1'(\theta)^2\right) - \frac{i}{4G\ell'}\int_0^{2\pi}d\theta\left(\{f_2(\theta),\theta\}-\frac{b^2}{2}f_2'(\theta)^2\right)}\,.
\eeq
Now, each Schwarzian action is invariant under $\theta \to \theta + \delta$, the rotational symmetry of each asymptotic region. Translations along the past circle are generated by the momentum stored in the past Schwarzian mode\footnote{We are normalizing the generators of rotations as one usually does in the Schwarzian theory, where we use $\ell$ to put the Schwarzian model on a circle with $\tau\sim \tau+|\ell|$, with $R$ generating the translation $\tau \to \tau + \varepsilon$.}
\beq
	R' = \frac{1}{4G\ell'^2}\left( \{f_2(\theta),\theta\} - \frac{b^2}{2}f_2'(\theta)^2\right)\,.
\eeq
There is a similar expression for the momentum $R$ of the future Schwarzian mode. $\hat{p}$ acts to the right on the normalized overlap as $i \frac{\partial}{\partial \ell'}$ (recall that we define $\hat{p}$ to satisfy minus the canonical commutator, $[\hat{\ell},\hat{p}]=-i$) and therefore acts on the unnormalized overlap as $i \sqrt{\ell'} \frac{\partial}{\partial \ell'}\frac{1}{\sqrt{\ell'}}$. (It acts to the left as $-i\sqrt{\ell}\frac{\partial}{\partial \ell}\frac{1}{\sqrt{\ell}}$.) Acting on~\eqref{E:1bdyoverlap}, we see that $\hat{p}$ inserts a factor of $R'-\frac{i}{2\ell'}$ into the JT path integral. Evidently quantum effects imply that $R'$ is not promoted to a Hermitian operator in the quantum theory, but the linear combination $R'-\frac{i}{2\ell'}$ is. We \emph{define} the operator $\hat{R}$ by this Hermitian combination so that $\hat{p} = \hat{R}$. The ``momentum'' canonically conjugate to $\hat{\ell}$ is in fact the momentum contained in the boundary Schwarzian mode.

With this result in hand we can understand the fact that negative momentum states have $\sim 0$ survival probability. Classically, we obtain a de Sitter space by gluing two asymptotic regions each characterized by the same $\alpha^2=-b^2$. These spaces are non-singular when $\alpha^2>0$, for which the spacetime has the same of a hyperboloid and the minimal length geodesic around the bottleneck has a length $2\pi |\alpha|$. However there are also singular spaces when $\alpha^2<0$, for which the spacetime has the shape of two cones with cone angles $2\pi |\alpha|$ glued together at the tips. These latter spaces are the JT analogue of crunching cosmologies, although JT gravity does not allow spacetime to simply end at the crunch, instead it inflates again to the future.

In the classical approximation the smooth geometries are characterized by equal positive momenta $\sim \alpha^2$ in the past and in the future, while the crunching geometries are characterized by equal negative momenta in the past and future. So semiclassically we expect for $\U$ to act unitarily on positive momentum states, and to annihilate negative momentum states, and indeed this is exactly what we find. What is surprising is that this is true beyond the semiclassical approximation, non-perturbatively as a function of $G$.

Let us take stock of the structure we have found so far.  We have Hilbert spaces of canonically isomorphic 1-boundary asymptotic states $\mathcal{H}_{P}^{1\text{-bdy}}$ and $\mathcal{H}_{F}^{1\text{-bdy}}$. The $\ket{\ell}$'s give an orthonormal basis on these spaces. Each Hilbert space is also furnished with canonically conjugate operators $\hat{\ell}$ and $\hat{p}$, which act like the usual position and momentum operators. $\hat{\ell}$ measures the signed length of the boundary, and $\hat{p}$ generates asymptotic rotations. Finally, $\widehat{\mathcal{U}}$ acts unitarily on the subspace of positive momentum states up to non-perturbative corrections.

\subsection{Many boundaries and Fock space}

Having discussed the $1 \to 1$ processes, we generalize our analysis to account for more boundaries.  In doing so, our quantum mechanics of single boundary states (analogous to single particle states) above will generalize to a Fock space generated by operators which create and destroy boundaries. The appearance of a Fock space suggests that we have a second quantized description, although it is in fact third quantized since creation operators create baby universes. However, in JT gravity the baby universes do not have additional fields living within them (i.e., gauge fields or matter fields), and so this third-quantized description is rather simple.

We have already constructed the isomorphic Hilbert spaces of asymptotic single-boundary states $\mathcal{H}_P^{1\text{-bdy}}\simeq \mathcal{H}_F^{1\text{-bdy}}\simeq \mathcal{H}^{1\text{-bdy}}$.  The Hilbert spaces of asymptotic 2-boundary states are 
\beq
	\mathcal{H}^{2\text{-bdy}} \simeq  \text{span}\{\ket{\ell_1,\,\ell_2} \}_{\ell_1,\,\ell_2 \in \mathbb{R}}\,, 
\eeq
where $\ket{\ell_1,\,\ell_2} = \ket{\ell_2,\,\ell_1}$, i.e.~asymptotic boundaries are identical bosons. Similarly,
\beq
	\mathcal{H}^{m\text{-bdy}} \simeq  \text{span}\{\ket{\ell_1,\ell_2,...,\ell_m}\}_{\ell_1,\ell_2,...,\ell_m \in \mathbb{R}}\,, 
\eeq
and these states are invariant under any permutation of the $\ell_i$'s.  Then in the genus expansion\footnote{With this qualification we are leaving open the possibility that the exact Hilbert space differs from the one we find in the genus expansion.} the full Hilbert space of asymptotic states is the Fock space
\beq
	\mathcal{H} \simeq \bigoplus_{m=1}^\infty \mathcal{H}^{m\text{-bdy}}\,.
\eeq
The same logic that went into the calculation of the inner product of 1-boundary states implies that, within the genus expansion, $m$-boundary states are orthogonal to $n$-boundary states unless $m=n$, and when $m=n$ we have up to doubly non-perturbative effects
\beq
\label{E:nUniverseIP}
	\braket{\ell_1,\ell_1,\hdots,\ell_m|\ell_1',\ell_2',\hdots,\ell_m'} \simeq \sqrt{\ell_1}\cdots \sqrt{\ell_m}\,\delta(\ell_1-\ell_1')\cdots \delta(\ell_m-\ell_m') + (\text{permutations}) \,.
\eeq
In this Subsection we work in the genus expansion, and so normalize these states via $\ket{\ell_1,\hdots ,\ell_m} \to \frac{\ket{\ell_1,\hdots,\ell_m}}{\sqrt{\ell_1}\hdots \sqrt{\ell_m}}$. This Fock space is isomorphic to that of a non-relativistic one-dimensional boson.

 It will be convenient to introduce creation and annihilation operators. We define a formal state $\ket{\Omega}$ satisfying $\braket{\Omega|\Omega}=1$ and $\braket{\Omega|\ell_1,\hdots\ell_m}=0$. We assign $\U\ket{\Omega}=\ket{\Omega}$, so that this state decouples from all physical states. Then define universe creation and annihilation operators $a_\ell^\dagger$ and $a_\ell$ with
\beq
\label{E:canonComm1}
	[a_\ell, a_{\ell'}^\dagger] = \delta(\ell - \ell')\,,\qquad [a_\ell, a_{\ell'}]  = [a_\ell^\dagger, a_{\ell'}^\dagger]  = 0\,.
\eeq
Then
\beq
	\ket{\ell_1,\ell_2,\hdots,\ell_m} = \frac{a_{\ell_1}^\dagger a_{\ell_2}^\dagger \cdots a_{\ell_m}^\dagger \ket{\Omega}}{\sqrt{\bra{\Omega} a_{\ell_m} \cdots a_{\ell_2} a_{\ell_1} a_{\ell_1}^\dagger a_{\ell_2}^\dagger \cdots a_{\ell_m}^\dagger \ket{\Omega}}}\,,
\eeq
and 
\beq
	\mathcal{H}^{m\text{-bdy}} \simeq  \text{span}\{a_{\ell_1}^\dagger a_{\ell_2}^\dagger \cdots a_{\ell_m}^\dagger\ket{\Omega} \}_{\ell_1,\ell_2,...,\ell_m \in \mathbb{R}}\,.
\eeq
We upgrade $\hat{\ell}$ from an operator that acts on 1-boundary states to an operator on the Fock space by defining
\beq
	\hat{\ell} = \int_{-\infty}^\infty d\ell \, \ell \, a_\ell^\dagger a_\ell\,.
\eeq
The $\ket{\ell_1,\hdots,\ell_m}$'s are eigenstates of $\hat{\ell}$ with
\beq
	\hat{\ell}\ket{\ell_1,\ell_2,\hdots,\ell_m} = (\ell_1 + \ell_2 + \cdots + \ell_m) \ket{\ell_1,\ell_2,\hdots,\ell_m}\,.
\eeq
We similarly upgrade $\hat{p}$, defining the operator which creates a boundary with momentum $p$ as 
\beq
	\tilde{a}_p^\dagger = \frac{1}{\sqrt{2\pi}} \int_{-\infty}^\infty d\ell \, e^{i p \ell} \, a_\ell^\dagger
\eeq
from which the standard commutation relations follow from~\eqref{E:canonComm1},
\beq
	[\tilde{a}_p, \tilde{a}_{p'}^\dagger] = \delta(p-p')\,,\qquad [\tilde{a}_p ,\tilde{a}_{p'}]  = [\tilde{a}_p^\dagger, \tilde{a}_{p'}^\dagger]  = 0\,,
\eeq
so
\beq
	\ket{p_1,p_2,...,p_m} = \frac{\tilde{a}_{p_1}^\dagger \tilde{a}_{p_2}^\dagger \cdots \tilde{a}_{p_m}^\dagger\ket{\Omega}}{\sqrt{\bra{\Omega} \tilde{a}_{p_m} \cdots \tilde{a}_{p_2} \tilde{a}_{p_1} \tilde{a}_{p_1}^\dagger \tilde{a}_{p_2}^\dagger \cdots \tilde{a}_{p_m}^\dagger \ket{\Omega}}}\,.
\eeq
Then 
\begin{equation}
\hat{p} = \int_{-\infty}^\infty dp \, p \, \tilde{a}_p^\dagger \tilde{a}_p\,,
\end{equation}
and
\begin{equation}
\hat{p}\ket{p_1,p_2,...,p_m} = (p_1 + p_2 + \cdots + p_m) \ket{p_1,p_2,...,p_m}\,.
\end{equation}
With these upgraded definitions of $\hat{\ell}$ and $\hat{p}$, we find the commutation relation
\begin{equation}
[\hat{\ell}, \,\hat{p}] = - i \, \widehat{N}\,,
\end{equation}
where $\widehat{N} = \int_{-\infty}^\infty d\ell \, a_\ell^\dagger a_\ell = \int_{-\infty}^\infty dp \, \tilde{a}_p^\dagger \tilde{a}_p$ is the number operator counting the number of boundaries.

The geometric interpretation is that $\hat{\ell}$ measures the total signed length of the boundary, while $\hat{p}$ generates symmetrized asymptotic rotations, consistent with the boundaries' bosonic statistics.

Having understood the natural operators acting on $\mathcal{H}$ we 
consider the transition amplitudes between normalized multi-boundary states,
\beq
	\bra{ \ell_1, \ell_2,...,\ell_n}\U\ket{\ell_1',\ell_2',...,\ell_{m}'}\simeq  \frac{Z_{n,m}(\ell_i,\ell'_j)}{\sqrt{\ell_1}\cdots \sqrt{\ell_n}\sqrt{\ell'_1}\cdots \sqrt{\ell_m'}}\,.
\eeq
Recall that $Z_{n,m}$ is the formal sum over surfaces which connect the $m$ past circles to the $n$ future circles.
These amplitudes approximately factorize in the genus expansion. The leading behavior when $m\neq n$ is suppressed by $|n-m|$ powers of $e^{-S_0}$, but when $n=m$, the amplitude is $O(1)$. That leading order result comes from ``Wick contractions'' in which one sums over every possible pairing between past and future boundaries, and each pairing is weighted by the ``free propagator'' of a single universe~\eqref{eq:normalizedprop2}. See Fig.~\ref{fig:2to2process}. In momentum space 
\beq
\label{eq:Uleadingorder}
	\bra{ p_1, p_2,...,p_n}\U\ket{p_1', p_2', ..., p_n' }\simeq \Theta(p_1)\cdots \Theta(p_n) \delta(p_1-p_1')\cdots \delta(p_n-p_n') + (\text{contractions}) + O(e^{-2S_0})\,.
\eeq
More succinctly, $\U$ acts unitarily (as the identity) on the Fock space of states where each momentum is positive, and annihilates the remaining states up to non-perturbative corrections.

\subsection{Non-perturbative corrections}
\label{S:Ugenus}

\textbf{Historical note added in [v3]:} The non-perturbative analysis below is contingent on a conjecture for the analytic continuation of Weil-Petersson volumes from~\cite{Cotler:2019nbi}, which was later disproved in~\cite{Turiaci:2020fjj, Eberhardt:2023rzz}.  Accordingly, the non-perturbative corrections and ensuing matrix model below do not hold.  However, a careful treatment of the nearly-dS$_2$ JT gravity path integral, including its measure, was performed in~\cite{Cotler:2024xzz}, leading to a new matrix model which does not depend on any conjectures. The measure appearing in that work is slightly different than what is described in this work, and explains the presence of a quantum mechanical interpretation with right-sign inner products. The new analysis in~\cite{Cotler:2024xzz} has similar formulae and conclusions as the analysis below.
\\ \\
We now undertake the computation of the genus corrections to~\eqref{eq:Uleadingorder}. Using our previous work~\cite{Cotler:2019nbi}, the genus $g$ correction to the normalized transition amplitudes $\bra{\ell_1,\hdots,\ell_n}\U\ket{\ell_1',\hdots,\ell_m'}$ takes the form
\beq
\label{eq:genuscorrection1}
	\int_0^{\infty} \prod_{i=1}^n \frac{db_i^2}{\ell_i} \prod_{j=1}^m \frac{db_j'^2}{\ell_j'}\, P_{g,n,m}(b_i^2,b_j'^2) \exp\left( -\frac{i \pi}{4G}\sum_{i=1}^n \frac{b_i^2}{\ell_i} + \frac{i \pi}{4G}\sum_{j=1}^m \frac{b_j'^2}{\ell'_j}\right)\,,
\eeq
where $P_{g,n,m}$ is a polynomial which depends on the number of boundaries $n$ and $m$, and the genus $g$. The result obtained by integrating Eqn.~\eqref{eq:genuscorrection1} is itself a polynomial in the $\ell_i$ and $\ell_j'$. So the corresponding momentum-space amplitudes are distributions supported at zero momentum. It follows that, on any superposition of states $\ket{p_1,\hdots,p_n}$ with all $p_i$ outside of an open set containing zero, $\U$ acts as $\Theta(p_i)$ to all orders in the genus expansion.

The genus corrections nevertheless lead to interesting effects. The $O(e^{-S_0})$ contribution to $\U$ comes from genus $0$ surfaces with three boundaries, either two past and one future or one past and two future. For the latter, we find a normalized transition amplitude
\beq
	\bra{\ell_1,\ell_2}\U\ket{\ell'} \simeq \left(\frac{\sqrt{\pi}}{2G}\right)^3\frac{e^{-S_0}}{\sqrt{-i}} \int_0^{\infty} \frac{db_1^2}{\ell_1}\frac{db_2^2}{\ell_2}\frac{ db'^2}{\ell'} \exp\left( -\frac{i \pi}{4G}\left(\frac{b_1^2}{\ell_1}+\frac{b_2^2}{\ell_2}-\frac{b'^2}{\ell'}\right) \right)+O(e^{-3S_0})\,,
\eeq
which leads to the momentum-space amplitude
\beq
	\bra{p_1,p_2}\U\ket{p'} \simeq 2^{3/2}\sqrt{-i} \,e^{-S_0}\delta(p_1)\delta(p_2)\delta(p') + O(e^{-3S_0})\,.
\eeq
This implies
\beq
\label{E:UNLO}
	\U \simeq\Theta(\hat{p}_i) + 2 e^{-S_0}\left( \sqrt{-i}\, \tilde{a}^{\dagger \,2}_0\tilde{a}_0 + \text{h.c.}\right) + O(e^{-2S_0})\,,
\eeq
where $\Theta(\hat{p}_i)\ket{p_1,\hdots p_n} = \Theta(p_1)\cdots \Theta(p_n)\ket{p_1,\hdots,p_n}$. 

The next correction of $O(e^{-2S_0})$ comes from (i) genus-1 surfaces with two boundaries, and (ii) genus-0 surfaces with four boundaries. These imply corrections to momentum-space amplitudes supported at $p=0$, although these corrections involve derivatives of delta functions. 

Consider a general state $\ket{\Psi}$ characterized by $n$-universe wavefunctions $\psi_n(p_1,\hdots,p_n)$ with
\beq
	\ket{\Psi} = \sum_{n=1}^{\infty} \frac{1}{n!}\int \prod_{i=1}^n dp_i \, \psi_n(p_1,\hdots,p_n) \ket{p_1,\hdots,p_n}\,,
\eeq
normalized as
\beq
	1 = \braket{\Psi|\Psi}= \sum_{n=1}^{\infty}\int \prod_{i=1}^ndp_i \,|\psi_n(p_1,\hdots,p_n)|^2\,.
\eeq
We take the wavefunctions $\psi_n$ to be symmetric in all arguments. One measure of the violation of unitarity is the deviation of $\bra{\Psi}\U^{\dagger}\U\ket{\Psi}=\big\|\,\U\ket{\Psi}\!\big\|_{L^2}^2$ from unity, which measures how the norm of $\ket{\Psi}$ evolves from the infinite past to the infinite future. Using~\eqref{E:UNLO} we find
\begin{align}
	\bra{\Psi}\U^{\dagger}\U\ket{\Psi} &\simeq \sum_{n=1}^{\infty}\left( \int_0^{\infty} \prod_{j=1}^n dp_j\,|\psi_n(p_1,\hdots,p_n)|^2+2^{3/2}e^{-S_0}n(n+1)\times \right.
	\\
	\nonumber
	& \qquad \left.\int_0^{\infty} \prod_{j=1}^{n-1}dp_j \left( \sqrt{-i}\,\psi_{n+1}^*(0,0,p_1,\hdots,p_{n-1})\psi_n(0,p_1,\hdots,p_{n-1}) + \text{h.c.}\right)+O(e^{-2S_0})\right)\,.
\end{align}
For wavefunctions supported over momenta $p\geq p_0>0$ for some fixed $p_0$ one has $\bra{\Psi}\U^{\dagger}\U\ket{\Psi}=1$, while wavefunctions that are supported in a neighborhood of zero already have $\bra{\Psi}\U^{\dagger}\U\ket{\Psi}=\big\|\,\U\ket{\Psi}\!\big\|_{L^2}^2<1$  on account of the support at negative momentum. The $O(e^{-S_0})$ correction is sign-indefinite.

In between these two cases are wavefunctions which are zero for negative momentum, but whose derivatives are nonzero at $p=0$. For such states the leading approximation to the matrix element is $\bra{\Psi}\U^{\dagger}\U\ket{\Psi}\simeq 1+O(e^{-2S_0})$, with the $O(e^{-S_0})$ term vanishing, and the corrections are sensitive to the derivatives of the $\psi_n$ at $p=0$. However, we expect that such states lie outside the regime of validity of the genus expansion. As we explain in Subsection~\ref{S:MMtoAmplitudes}, the leading doubly non-perturbative correction is a rapidly oscillatory term $\sim e^{ie^{S_0}}$ which effectively smears wavefunctions over a momentum scale $e^{-S_0}$. So we expect that our predictions from the genus expansion are only reliable when they  give the same result as for a smeared wavefunction.

Taken together, we see that within the regime of validity of our computations we have $\bra{\Psi}\U^{\dagger}\U\ket{\Psi}=\big\|\,\U\ket{\Psi}\!\big\|_{L^2}^2 \leq 1$, suggesting that we can think of de Sitter JT gravity as being an open quantum system. It would be interesting to study the fate of those ``borderline'' states from the last paragraph, by properly accounting for rapidly oscillating corrections.  Below, when we study the matrix integral dual of de Sitter JT gravity, we will find evidence that these corrections are associated with tunneling between states with positive momenta and states with negative momenta.

\subsection{A few comments on horizon entropy}
\label{S:entropy}

We conclude this Section with some comments on the horizon entropy of nearly dS$_2$ space. The classical approximation to the horizon entropy of nearly dS$_2$ JT gravity is $S_{\rm cosmo}\approx 2 S_0$\,. The horizon entropy is often taken as motivation for the claim that the Hilbert space of de Sitter quantum gravity is finite-dimensional, with $\text{dim}(\mathcal{H}) \sim e^{S_{\rm cosmo}}\sim e^{2S_0}$. Above, we saw that the Hilbert space of asymptotic states was infinite dimensional: we may have an arbitrarily large number of boundaries, each labeled by an independent $\ell$. These two comments are not necessarily in conflict with each other. After all, the Hilbert space of asymptotic states described above is only \text{defined} in the genus expansion, with $e^{-S_0}\ll 1$ playing the role of the small expansion parameter. This is consistent with a Hilbert space dimension of order $\sim e^{2S_0}$, which would be non-perturbative in the genus expansion.  Further, to the extent that the horizon entropy is counting states, it is not clear if it is computing asymptotic or bulk states.  While there is an expectation that the entropy is counting bulk states, distinguishable over a de Sitter time, presently we only know how to count asymptotic states. Clearly, to assess whether the Hilbert space of asymptotic states is finite- or infinite-dimensional, we require a non-perturbative formulation of de Sitter JT gravity. Fortunately, we have such a formulation in the dual matrix integral, and we discuss this question further in Section~\ref{S:MM}.

Given the simplicity of JT gravity it would be interesting to compute the horizon entropy beyond the semiclassical approximation. Perhaps one can make sense of the entanglement entropy in JT for half of space, along the lines of the analysis~\cite{Lin:2018xkj}. We leave this question for future work.

\section{Matrix model interpretation}
\label{S:MM}

In the last Section we showed how to extract the Hilbert space of asymptotic states and transition amplitudes from the de Sitter JT path integral. There is significant evidence~\cite{Maldacena:2019cbz,Cotler:2019nbi,Saad:2019lba} that de Sitter JT gravity is dual to a matrix integral. In this Section we show how the Hilbert space and approximately unitary evolution 
emerge from the matrix side of the duality. Furthermore, this 
emergence is robust in a large class of matrix models, arising from universal features of nearest-neighbor eigenvalue repulsion in a certain large matrix limit.
The random matrix description of de Sitter JT gravity is particularly striking, since a 1+1 dimensional quantum theory is arising from a 0+0 dimensional classical integral.

\subsection{Matrix models and the holographic dictionary}

We first review the connection between JT gravity and matrix integrals, as well as matrix integrals more broadly.  Letting $H$ be a $d \times d$ Hermitian matrix and $f(H)$ be a multi-trace function thereof, we can consider a large-$d$ matrix model whose expectation values are computed by
\beq
\label{E:MMaverage}
	\langle f(H) \rangle_{\rm MM} = \frac{1}{\mathcal{Z}} \int dH \, e^{- d \, \text{tr}(V(H,d))} f(H)
\eeq
where $V(H,d)$ is a power series $H$ and $\mathcal{Z} = \int dH\, e^{-d \,\text{tr}(V(H))}$ is the matrix partition function. We denote matrix averages with $\langle \cdot \rangle_{\rm MM}$ to distinguish the average from the quantum mechanical average for JT gravity in the last Section. The measure $dH$, potential $\text{tr}(V(H,d))$, and observable $f(H)$ are invariant under $H \to U H U^\dagger$ for any $U \in U(d)$, which may be used to write the matrix average as an integral over eigenvalues.

Let $\{\lambda_1,...,\lambda_d \}$ be the eigenvalues of a matrix $H$.  We denote the leading normalized density of states in the large-$d$ limit as
\beq
\rho_0(E) = \lim_{d \to \infty}\left\langle \frac{1}{d} \sum_{j=1}^d \delta(E - \lambda_j) \right\rangle_{\rm MM} \,,
\eeq
which obeys $\int dE \, \rho_0(E) = 1$.  The function $\rho_0(E)$ is defined on the real line, and for most sensible potentials its support is the union of a finite number of disjoint intervals. If the support of $\rho_0(E)$ is a single interval $[a,b]$, then we say the matrix integral is 1-cut. Beyond leading order, there is a genus expansion in the parameter $1/d$ for the exact density of states $\rho(E)$ and likewise for other matrix averages.

As an example, consider
\beq
	V(H) = \frac{8H^2}{c^2}\,,
\eeq
which has a single cut $\left[-\frac{c}{2},\frac{c}{2}\right]$ with a leading density of states
\beq
	\rho_0(E) = \frac{8}{\pi c^2}\sqrt{E^2-\frac{c^2}{4}}\,.
\eeq
We are interested in the ``double-scaled'' limit. Rather than first giving a precise definition, let us explain how to obtain it for the quadratic model. First, let us shift the potential $V(H) \to \frac{8}{c^2}\left(H-\frac{c}{2}\right)^2$, so that the density of states becomes $\rho_0(E) \to \frac{8}{\pi c^2}\sqrt{E(c-E)}$ and the cut is shifted to $[0,c]$. Next define the total density of states $\rho_0^{\rm tot}(E)=d\,\rho_0$, which satisfies $\int dE \,\rho_0^{\rm tot}(E)=d$. Now take the double-scaling limit, in which we send both $d$ and $c$ to infinity while keeping the ratio $\frac{d}{c^{3/2}} \equiv e^{S_0}\gg 1$ fixed. This limit results in a finite total density of states,
\beq
\label{E:GUE}
	\rho_0^{\rm tot}(E) = \frac{8 \,e^{S_0}}{\pi}\sqrt{E}\,,
\eeq
so there is a single cut $[0,\infty)$ with a finite but large density of states. In the double-scaled model there is still a genus expansion for the exact density of states and for matrix averages, where now the genus expansion parameter is $e^{-S_0}$. 

For a more general model the double-scaling limit is achieved by a similar series of steps. We take a one-cut model, shift the potential so that the cut lies on $[0,c]$, and then simultaneously take $d$ and $c$ to infinity with $d\gg c$. The precise limit depends on the details of the potential, but it should be taken in such a way that the limit of the total density of states is large and finite, with $\rho_0^{\rm tot}(E)$ proportional to a parameter $ e^{S_0}\gg 1$. The resulting double-scaled model has a perturbative expansion in $e^{-S_0}$. 

Ordinary large $d$ matrix integrals enjoy remarkable properties. For us, the most important of them is the $1/d$ expansion of matrix averages is determined by $\rho_0(E)$ through a procedure known as topological recursion~\cite{Eynard:2007kz}. See~\cite{Saad:2019lba} for a practical introduction and~\cite{Eynard:2015aea} for a review.  This procedure commutes with the double scaling limit in understood examples, with the $1/d$ expansion being replaced by an $e^{-S_0}$ expansion. 

\begin{figure}
\begin{center}
\includegraphics[width=6in]{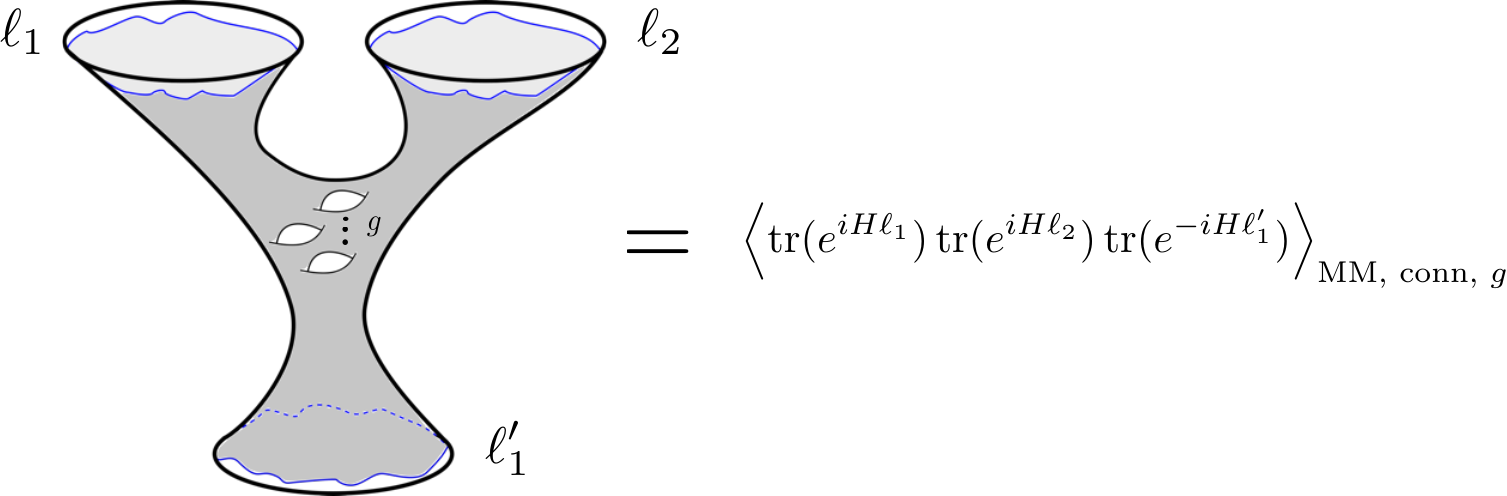}
\end{center}
\caption{\label{F:dictionary} An entry in the dictionary between the matrix integral and de Sitter JT gravity. The JT partition function on a genus $g$ surface with two future boundaries and one past boundary is equal to the indicated three-point function of the matrix ensemble. See~\eqref{E:dictionary}.}
\end{figure}

It has recently been established~\cite{Saad:2019lba} that JT gravity in Euclidean AdS$_2$ is equivalent to a double-scaled matrix integral to all orders in the genus expansion. Building upon that result, significant evidence was presented in~\cite{Maldacena:2019cbz,Cotler:2019nbi} that de Sitter JT gravity is also equivalent to a double-scaled matrix integral, again to all orders in the genus expansion. The basic dictionary has two parts. First, the matrix integral is specified by the leading order density of states
\beq
\label{E:JTdos}
	\rho_0^{\rm tot}(E) = \frac{\sqrt{G}}{2\pi^{3/2}}\,e^{S_0}\sinh\left(\sqrt{\frac{\pi E}{G}}\right)\,,
\eeq
where $G$ is the JT gravitational coupling and $e^{-S_0}$ its genus expansion parameter. The $e^{-S_0}$ expansion of matrix observables is then determined by this density of states via topological recursion. Second, a future boundary with signed length $\ell$ corresponds to an insertion of the operator 
\beq
\label{E:bdyInsertion}
	\text{tr}\left( e^{i \ell H}\right) = \int_0^{\infty} dE\, \rho(E) \, e^{i \ell E}\,,
\eeq
where $\rho(E)=\rho^{\rm tot}(E) + O(e^{-S_0})$ is the exact density of states, along with an $i\epsilon$ prescription whereby $\ell$ has small positive imaginary part. A past boundary with signed length $\ell$ corresponds to the complex conjugate insertion $\text{tr}\left( e^{-i \ell H}\right)$, where $\ell$ has small negative imaginary part. Then the holographic dictionary between gravity and matrix observables is as follows. Recall that $Z_{g,n,m}(\ell_i,\ell_j')$ is the JT partition function on a genus $g$ surface with $n$ future boundaries of signed lengths $\ell_i$ and $m$ past boundaries of signed lengths $\ell_j'$. Stripping off its $e^{S_0}$-dependence produces the expansion coefficients $\Psi_{g,n,m}(\ell_i,\ell_j')$, which equal
\beq
\label{E:dictionary}
	\Psi_{g,n,m}(\ell_i,\ell_j') = \left\langle \prod_{i=1}^n\text{tr}\left( e^{i \ell_iH} \right)\prod_{j=1}^m\text{tr}\left(e^{-i\ell'_jH}\right)\right\rangle_{\text{MM, conn},\,g}\,.
\eeq
The right-hand-side is the average in the double-scaled matrix model with leading density of states~\eqref{E:JTdos}. The subscript ``conn'' indicates that we are looking at the connected part of this matrix correlation function, and $g$ that it is the genus $g$ term in its genus expansion. See Fig.~\ref{F:dictionary} for an example. As another example, consider the disk partition function with a single future circle, $Z_{0,1,0}=e^{S_0}\,\Psi_{0,1,0}$, given in~\eqref{E:diskZ}. It gave the genus-0 approximation to the unnormalized Hartle-Hawking wavefunction. According to this dictionary and using the leading order density of states~\eqref{E:JTdos}, the disk partition function is $e^{S_0}$ times
\beq
	\left\langle \text{tr}\left( e^{i \ell H}\right)\right\rangle_{\text{MM, conn},\,0} = \int_0^{\infty} dE \, \frac{\sqrt{G} }{8\pi^{3/2}} \sinh\left(\sqrt{\frac{\pi E}{G}}\right) e^{i \ell E} = \frac{1}{\sqrt{2\pi}(-2 i \ell)^{3/2}}\,e^{\frac{\pi i}{4G \ell}}\,,
\eeq
which matches our result from JT gravity~\eqref{E:diskZ}.

In our previous work~\cite{Cotler:2019nbi}, we largely focused our attention on the genus expansion coefficients $\Psi_{g,n,m}$. In this article we have used these coefficients to uncover the transition amplitudes of de Sitter JT gravity, as well as the quantum mechanics underlying them. Below we will formulate a more refined dictionary relating these transition amplitudes and other features of the quantum mechanics of the de Sitter JT gravity to the matrix integral side of the duality.

\subsection{Amplitudes from matrix integrals
\label{S:MMtoAmplitudes}}

Mirroring our discussion in Section~\ref{S:bulkH}, we begin with the matrix model description of the $1\to 1$ processes, and then discuss the multi-boundary processes.

\subsubsection{Single-boundary transitions}

Consider the unnormalized transition amplitude in JT gravity between a past circle with signed length $\ell'$ and a future circle with signed length $\ell$, $\bra{\ell}\U\ket{\ell'}$. It is a sum over surfaces with one past and one future boundary, and so 
\beq
	\bra{\ell}\U\ket{\ell'} \simeq Z_{1,1}(\ell,\ell') = \sum_{g=0}^{\infty} e^{-2g S_0}\,\Psi_{0,1,1}(\ell,\ell')\,.
\eeq
We write a $\simeq$ because the genus expansion is asymptotic. From our review, this matrix element has the same genus expansion as an average in the dual matrix model,
\beq
\label{E:1to1Dictionary}
	\bra{\ell}\U\ket{\ell'} \simeq \left\langle \text{tr}\left(e^{i\ell H}\right)\text{tr}\left( e^{-i \ell'H}\right)\right\rangle_{\rm MM,\, conn}\,.
\eeq	
Beyond the genus expansion, we \emph{define} the amplitude by this matrix average, since the latter is non-perturbatively well-defined (although this is subtle for the JT matrix model; see~\cite{Saad:2019lba}). Note that the $1\to 1$ amplitude is simply the spectral form factor of the dual matrix ensemble.

In Section~\ref{S:bulkH} we showed how to compute the inner product of 1-boundary states using the JT path integral, with the result $\braket{\ell|\ell'} \simeq \sqrt{\ell}\sqrt{\ell'} \,\delta(\ell-\ell')$ up to non-perturbative corrections in $e^{-S_0}$. The dual matrix model does not have a Hilbert space, and so it is not necessarily obvious how to establish a dictionary between this inner product and an observable in the matrix model. Our approach is to reverse-engineer the translation protocol. We note that the genus expansion of this amplitude in JT gravity has the form
\beq
	Z_{1,1}(\ell,\ell') = \frac{i}{2\pi}\frac{\sqrt{\ell}\sqrt{\ell'}}{\ell-\ell'+i\epsilon} + (\text{regular in }\ell,\ell')\,.
\eeq
The expansion coefficients have square root branch cuts\footnote{We take the branch cuts to be along the rays $\text{arg}(\ell) =-\pi/2$ and $\text{arg}(\ell')=\pi/2$. This choice may seem arbitrary, but it is what follows from the analytic continuation of $Z_{1,1}$ from JT gravity in Euclidean AdS$_2$, or from the two-point function of $\text{tr}\left(e^{-\beta H}\right)$ in the matrix integral. In either case, $\ell$ and $\ell'$ are effectively the analytic continuations of an inverse temperature $\beta \to -i \ell$, $\beta'\to i \ell'$, where the branch cuts are originally along the ray $\beta,\beta' <0$.} but the genus-0 term shown above is the unique one with a singularity. That pole implies a singular discontinuity as we dial the imaginary part of $\ell$ from a
negative sign (which is unphysical) to a positive sign. That discontinuity is precisely the inner product we obtained from the JT path integral,
\beq
	\text{disc}\,Z_{1,1}(\ell,\ell') = \sqrt{\ell}\sqrt{\ell'}\,\delta(\ell-\ell')\,.
\eeq
Since $Z_{1,1}$ coincides with the spectral form factor to all orders in the genus expansion, the spectral form factor has the same singular discontinuity. We therefore define the dictionary for the inner product and the matrix model to be
\beq
\label{E:1to1MM}
	\braket{\ell|\ell'} = \text{disc}\,\left\langle \text{tr}\left( e^{i\ell H}\right)\text{tr}\left(e^{-i\ell'H}\right)\right\rangle_{\rm MM,\,conn}\,.
\eeq

Neglecting non-perturbative effects in $e^{-S_0}$, the transition amplitude between normalized states $\bra{\ell}$ and $\ket{\ell'}$ is 
\beq
\label{E:spectralFF}
	\bra{\ell}\U\ket{\ell'} \simeq \frac{Z_{1,1}(\ell,\ell)}{\sqrt{\ell}\sqrt{\ell'}}\simeq \left\langle \text{tr}\left(\frac{e^{i \ell H}}{\sqrt{\ell}}\right)\text{tr}\left( \frac{e^{-i \ell'H}}{\sqrt{\ell'}}\right)\right\rangle_{\text{MM,\,conn}}\,.
\eeq
Now recall that on the Hilbert space of 1-boundary states we had the signed length operator $\hat{\ell}$ and the momentum $\hat{p}$ (with the convention $[\hat{\ell},\hat{p}]=-i$).
Inserting $\hat{\ell}$ to the right of $\U$ simply produces a factor of $\ell'$ in the matrix average, while inserting it to the left inserts a factor of $\ell$. The momentum operator is more interesting. Inserting it to the right of $\U$, it acts on the normalized amplitude as $i \frac{\partial}{\partial \ell'}$, so that
\beq
	\bra{\ell}\U\,\hat{p}\ket{\ell'} \simeq \left\langle \text{tr}\left(\frac{e^{i\ell H}}{\sqrt{\ell}} \right)\text{tr}\left( \left( H-\frac{i}{2\ell'}\right)\frac{e^{-i\ell'H}}{\sqrt{\ell'}}\right)\right\rangle_{\rm MM,\,conn}\,.
\eeq
So, up to an additive constant, the momentum operator $\hat{p}$ corresponds to additional insertions of the matrix $H$ of the dual matrix integral. Similarly, an insertion of $\hat{p}$ to the left of $\U$ acts on the amplitude as $-i \frac{\partial}{\partial \ell}$. This modifies the ``future'' insertion in the matrix integral to be $\text{tr}\left(\left(H + \frac{i}{2\ell}\right)\frac{e^{i\ell H}}{\sqrt{\ell}}\right)$. 

The evolution operator was most naturally expressed in the momentum basis, with
\begin{equation*}
	\bra{p}\U\ket{p'} = \int_{\mathbb{R}+i \epsilon}\frac{d\ell}{\sqrt{2\pi}}\int_{\mathbb{R}-i\epsilon}\frac{d\ell'}{\sqrt{2\pi}} \, e^{-i p \ell+i p'\ell'}\bra{\ell}\U\ket{\ell'} \simeq \Theta(p)\delta(p-p') + O(e^{-2S_0})\,.
\end{equation*}
Using~\eqref{E:bdyInsertion}, we map this to the matrix average
\beq
	\bra{p}\U\ket{p'} \simeq \left\langle \text{tr}\left( \sqrt{\frac{2}{-i(H-p)}}\,\right)\text{tr}\left(\sqrt{\frac{2}{i(H-p')}}\,\right)\right\rangle_{\rm MM,\,conn}\,.
\eeq
In other words, within the genus expansion, a future boundary with momentum $p$ inserts what we term a \emph{half-resolvent} into the matrix average
\beq
	W_{\frac{1}{2}}(p) = \text{tr}\left(\sqrt{\frac{2}{-i(H-p)}}\,\right) = \sqrt{2}\int_0^{\infty} \frac{dE\,\rho(E)}{\sqrt{-i(E-p)}}\,,
\eeq
and a past boundary with momentum $p$ its complex conjugate.

Our result for the amplitude $\bra{p}\U\ket{p'} \simeq \Theta(p)\delta(p-p') + O(e^{-2S_0})$ is equivalent to the statement in the matrix model that the genus-0 approximation to the two-point function of half-resolvents is
\beq
	\left\langle W_{\frac{1}{2}}(p)W^*_{\frac{1}{2}}(p')\right\rangle_{\text{MM,\,conn},\,0} = \Theta(p)\delta(p-p')\,.
\eeq
We will have more to say about this in Subsection~\ref{S:MMgeneral}.

In JT gravity we understood the physics behind $\bra{p}\U\ket{p'}\sim \Theta(p)\delta(p-p')$ being supported over positive $p$.  Namely, semiclassically, smooth de Sitter geometries are characterized by positive past and future momenta, while negative momentum initial states end in a crunch. But in the dual matrix model the domain $p>0$ also has a clear interpretation: we have $\hat{p}\sim H$, and the subspace $p>0$ just corresponds to the semi-infinite cut $E> 0$ of the matrix model.

We also saw in Subsection~\ref{S:Ugenus} that the genus corrections to the amplitude lead to distributional corrections to $\bra{p}\U\ket{p'}$ supported at $p,p'=0$. One expects that these corrections build up a doubly non-perturbative effect near zero momentum. There is a natural guess for this effect on the matrix integral side. Namely, the leading doubly non-perturbative corrections near the end of the cut. These come in one of two types. While to all orders in the genus expansion eigenvalues are found with certainty within the cut $E\in [0,\infty)$, there is a non-perturbatively small likelihood $\sim e^{-e^{S_0}}$ for them to be found along the negative real axis. This will lead to a small $\sim e^{-e^{S_0}}$ value for the transition amplitude between positive and negative momenta, which is zero to all orders in the genus expansion. So these non-perturbative corrections will lead to tunneling from positive momentum to negative momentum in de Sitter. This can be regarded as a (doubly) non-perturbative instability. Within the cut, there are also rapidly oscillatory contributions $\sim e^{i e^{S_0}}$ to correlation functions of the density of states, which will in turn contribute rapid oscillations to the two-point function of half-resolvents near zero momentum.

\subsubsection{Multi-boundary transitions}

Now consider  transition amplitudes between unnormalized states with at least one boundary,
\beq
	\bra{\ell_1,\hdots,\ell_n}\U\ket{\ell_1',\hdots,\ell_m'} \simeq Z_{n,m}(\ell_i,\ell_i') \,.
\eeq
These amplitudes involve a sum over geometries which connect $m$ past circles of signed lengths $\ell_j'$ to $n$ future circles of signed lengths $\ell_i$, in which we project out the no-boundary state. When there is more than one universe in the future and in the past, the amplitude depends on disconnected geometries. See the text near Fig.~\ref{fig:2to2process} and Eq.~\eqref{E:2to2} for a discussion of the $2\to 2$ process. We would like to find the matrix average whose genus expansion coincides with this sum. 

To find it we consider the generating functional of connected correlation functions of future and past boundary operators. It is
\beq
	W(\zeta_+,\zeta_-) \equiv \ln \left\langle \exp\left( \int_{\mathbb{R}+i\epsilon} d\ell \,\zeta_+(\ell)\,\text{tr}\left( e^{i \ell H}\right)\right)\exp\left( \int_{\mathbb{R}-i\epsilon} d\ell'\,\zeta_-(\ell')\,\text{tr}\left(e^{-i\ell'H}\right)\right)\right\rangle_{\rm MM}\,.
\eeq
A variational derivative with respect to $\zeta_+(\ell)$ inserts a future boundary of signed length $\ell$, while a variational derivative with respect to $\zeta_-(\ell')$ inserts a past boundary of signed length $\ell'$. For example,
\beq
	\left\langle \text{tr}\left( e^{i\ell H}\right)\text{tr}\left(e^{-i\ell'H}\right)\right\rangle_{\rm MM, conn} = \left.\frac{\delta^2W}{\delta\zeta_+(\ell)\delta \zeta_-(\ell')}\right|_{\zeta_{\pm}=0}\,.
\eeq
The generating functional of correlation functions, including the disconnected parts, is $Z\equiv e^{W}$. We want to subtract the terms from this which correspond to bulk geometries in which the past is not connected to the future, for example, a disk anchored on a future circle. To do this we construct the generating functional of connected correlation functions with only future or only past boundaries,
\beq
	W_{\pm}(\zeta_{\pm}) \equiv \ln \left\langle \exp\left(\int_{\mathbb{R}\pm i\epsilon}d\ell\,\zeta_{\pm}(\ell)\,\text{tr}\left(e^{\pm i \ell H}\right)\right)\right\rangle_{\rm MM}\,,
\eeq
i.e. $W_+(\zeta_+)=W(\zeta_+,0)$ and $W_-(\zeta_-)=W(0,\zeta_-)$. Then the generating functional with only future or past boundaries is $Z_{\pm}(\zeta_{\pm})\equiv e^{W_{\pm}(\zeta_{\pm})}$. With a bit of thought, one realizes that subtracting off the geometries in which the past is disconnected from the future amounts to considering the ratio
\begin{align}
\begin{split}
	Z_{\rm dS}(\zeta_+,\zeta_-) &\equiv \frac{Z(\zeta_+,\zeta_-)}{Z_+(\zeta_+)Z_-(\zeta_-)} 
	\\
	&= \frac{\left\langle \exp\left(\int_{\mathbb{R}+i\epsilon}d\ell\,\zeta_+(\ell)\text{tr}\left(e^{i\ell H}\right)\right)\exp\left(\int_{\mathbb{R}-i\epsilon}d\ell'\,\zeta_-(\ell')\text{tr}\left(e^{-i\ell'H}\right)
	\right)\right\rangle_{\rm MM}}{\left\langle \exp\left(\int_{\mathbb{R}+i\epsilon}d\ell\,\zeta_+(\ell)\text{tr}\left(e^{i\ell H}\right)\right)\right\rangle_{\rm MM}\left\langle \exp\left(\int_{\mathbb{R}-i\epsilon}d\ell'\,\zeta_-(\ell')\text{tr}\left(e^{-i\ell'H}\right)\right)\right\rangle_{\rm MM}}\,.
\end{split}
\end{align}
We term $Z_{\rm dS}$ the de Sitter generating functional.

We conclude that the genus expansion of the unnormalized amplitude $\bra{\ell_1,\hdots,\ell_n}\U\ket{\ell_1',\hdots \ell_m'}$ is given by $n$ derivatives of $Z_{\rm dS}$ with respect to $\zeta_+$ and $m$ derivatives with respect to $\zeta_-$, and then setting $\zeta_{\pm}$ to vanish. So as in our treatment of the $1\to 1$ amplitudes, we \emph{define} the unnormalized amplitudes by these quantities in the matrix model.

As a simple check, consider the $1\to 1$ process from the last Subsection. According to this definition, we have
\begin{align}
\begin{split}
	\bra{\ell}\U\ket{\ell'} &= \frac{\delta^2Z_{\rm dS}}{\delta \zeta_+(\ell)\delta \zeta_-(\ell')} = \left\langle \text{tr}\left( e^{i \ell H}\right)\text{tr}\left(e^{-i\ell'H}\right)\right\rangle_{\rm MM} - \left\langle \text{tr}\left(e^{i\ell H}\right)\right\rangle_{\rm MM}\left\langle \text{tr}\left(e^{-i\ell'H}\right)\right\rangle_{\rm MM} 
	\\
	&= \left\langle \text{tr}\left(e^{i\ell H}\right)\text{tr}\left(e^{-i\ell'H}\right)\right\rangle_{\rm MM,\,conn}\,,
\end{split}
\end{align}
which indeed reproduces our prescription in~\eqref{E:1to1MM}. For amplitudes with more than 1 boundary our prescription does not simply reduce to the connected correlator.

Now we turn our attention to the inner product on multi-boundary states. Following our dictionary for the inner product on the 1-boundary states, we identify the inner product on multi-boundary states from the singular discontinuity in the amplitude as we dial the imaginary parts of the $\ell_i$ from negative and small to positive and small. To all orders in the genus expansion the singularities in the amplitudes $\bra{\ell_i}\U\ket{\ell_j'}$ only come from the $1\to 1$ process, which on the matrix model side  is the singularity in the leading order spectral form factor. As a result there is only a singular discontinuity when $n=m$ with the same result as we found in Eq.~\eqref{E:nUniverseIP}. By design this dictionary gives the same inner product we found from gravity.

So within the genus expansion we have a Fock space of states emerging from the matrix model. The bosonic nature of the boundaries follows from the dictionary and the fact that $m$ past boundaries correspond to the insertion of $\prod_{j=1}^m\text{tr}\big(e^{-i\ell_j'H}\big)$ and $n$ future boundaries to the insertion of $\prod_{i=1}^n\text{tr}\big(e^{i_i\ell H}\big)$. So the matrix average will be completely symmetric under permutations of the $\ell'_j$ and of the $\ell_i$.

It would be of course extremely interesting to uncover the effect of doubly non-perturbative corrections on the amplitudes as well as on the inner product.

As with our discussion of the $1\to 1$ amplitudes, we can obtain a dictionary between the momentum space amplitudes and the matrix model by renormalizing our states by factors of $1/\sqrt{\ell}$ and then Fourier transforming from $\ell$ to $p$. The momentum space amplitudes are again related to correlation functions of half-resolvents in the matrix model, with a future boundary of momentum $p$ corresponding to the half-resolvent $W_{\frac{1}{2}}(p)$ and a past boundary to its complex conjugate $W_{\frac{1}{2}}^*(p)$. 

\subsection{Beyond the JT matrix integral}
\label{S:MMgeneral}

Above we have seen how the Hilbert space of asymptotic states and the infinite-time evolution operator $\U$ emerge from the JT matrix integral. In this Subsection we show that this emergence is a robust consequence of eigenvalue repulsion and the double scaling limit, as well as provide evidence that the non-perturbative Hilbert space has infinite dimension.

The main object in our gravitational analysis was the annulus partition function $Z_{0,1,1}(\ell,\ell')$ in~\eqref{E:annulusZ}, which gave the leading approximation to the transition amplitude $\bra{\ell}\U\ket{\ell'}$ between unnormalized 1-universe states. From this amplitude we obtained the propagator of a single universe as well as the inner product on 1-universe states. The dictionary between gravitational observables and matrix integrals equates this unnormalized amplitude with the spectral form factor~\eqref{E:spectralFF} of the matrix model.

Now, it is a general feature of large-$d$ matrix integrals that observables like the density of states depend sensitively on the potential of the model. However, the connected double resolvent, which depends on the two-point function of eigenvalue densities, is a universal function of the cuts. Ultimately this universal form owes its existence to the repulsion of nearby eigenvalues. For a matrix model with a single symmetric cut $[-\frac{c}{2},\frac{c}{2}]$ the connected double resolvent is
\beq
\label{E:R02}
	\left\langle \text{tr}\left(\frac{1}{E_1-H}\right)\text{tr}\left(\frac{1}{E_2-H}\right)\right\rangle_{\rm MM,\,conn} \simeq \frac{1}{2(E_1-E_2)^2}\left(\frac{E_1E_2-\frac{c^2}{4}}{\sqrt{E_1^2-\frac{c^2}{4}}\sqrt{E_2^2-\frac{c^2}{4}}}-1\right) + O(d^{-2})\,,
\eeq
which for a double-scaled model with a cut $[0,\infty)$ becomes
\beq
	\left\langle \text{tr}\left(\frac{1}{E_1-H}\right)\text{tr}\left(\frac{1}{E_2-H}\right)\right\rangle_{\rm MM,\,conn} \simeq \frac{1}{4\sqrt{-E_1}\sqrt{-E_2}(\sqrt{-E_1}+\sqrt{-E_2})^2}+O(e^{-2S_0})\,.
\eeq
This universal form determines the leading approximation to the spectral form factor through a double inverse Laplace transform in $E_1$ and $E_2$. In other words, the universality of the double resolvent implies the universality of the spectral form factor.

We conclude that the most basic features of de Sitter JT gravity, namely its Hilbert space and the leading order evolution operator, are a consequence of the most basic features of the dual matrix integral, namely the repulsion of nearby eigenvalues along with the double scaling limit. 

With this result in hand we might wonder whether this duality is but one instance of a more general relation between matrix integrals and models of 1+1 de Sitter quantum gravity.  Above we considered Hermitian matrix integrals, which belong to the Gaussian Unitary Ensemble (GUE) symmetry class.  In particular, our matrix integrals have $U(d)$-invariant measures.  Beyond JT gravity, another example of a double-scaled GUE class matrix integral is the limit of the Hermitian quadratic matrix model with density of states~\eqref{E:GUE}. This model is already known to be dual to topological gravity (see~\cite{Witten:1990hr,kontsevich1992} and a more recent discussion~\cite{Dijkgraaf:2018vnm}).  Perhaps the de Sitter dictionary we presented for the JT matrix integral setting has an analogue in topological gravity.

Besides the GUE class of random matrices there are nine other symmetry classes in the Altland-Zirnbauer classification~\cite{Altland:1997zz}.  In each case, double-scaled matrix integrals satisfy 
\beq
\label{E:symprop1}
	\left\langle \text{tr}\left(e^{i \ell H}\right)  \text{tr}\left(e^{-i \ell' H}\right) \right\rangle_{\text{MM}'\text{, conn}} \simeq \mathcal{C}_{\text{MM}'} \,\frac{i}{2\pi} \frac{\sqrt{\ell}\sqrt{\ell'}}{\ell - \ell'+ i \epsilon} + O(e^{-2 S_0})\,,
\eeq
where $\text{MM}'$ denotes the matrix integral of choice, and $\mathcal{C}_{\text{MM}'}$ is a positive constant depending on the symmetry class (e.g. $\mathcal{C}_{\text{MM}'} = 1$ for the GUE).  These universal results reflect appropriate modifications of eigenvalue repulsion to accommodate for symmetry.  As such, we have the discontinuity
\beq
	\label{E:symmdisc1}
\text{disc} \, \left\langle \text{tr}\left(e^{i \ell H}\right)  \text{tr}\left(e^{-i \ell' H}\right) \right\rangle_{\text{MM}'\text{, conn}} \simeq \sqrt{\mathcal{C}_{\text{MM}'} \, \ell} \sqrt{\mathcal{C}_{\text{MM}'} \, \ell'} \, \delta(\ell - \ell')\,.
\eeq
It is tempting to suggest that there exist models of nearly-dS$_2$ gravity with other symmetry classes, with asymptotic states having the inner product suggested by~\eqref{E:symmdisc1}, and accordingly leading-order unitarity on positive-momentum states due to~\eqref{E:symprop1}.

Recently, Stanford and Witten have studied generalizations of JT gravity in Euclidean AdS~\cite{Stanford:2019vob} corresponding to the various symmetry classes in the Altland-Zirnbauer classification.  For each generalization there is a dual matrix integral in the corresponding symmetry class.  It is likely that these dualities have Lorentzian de Sitter counterparts, and the universality of the two-point function~\eqref{E:symprop1} suggests that in each case a Hilbert space and $\U$ emerge from the matrix integral by the same mechanism as we find above.

More generally, we can consider double-scaled matrix integrals such that the corresponding random matrix $H$ decomposes into a direct sum $H = \bigoplus_i H_i$ (i.e., it is block diagonal), where each $H_i$ may even belong to a distinct symmetry class. Then~\eqref{E:symprop1} and~\eqref{E:symprop1} would hold within each subspace. We expect that this structure is realized, for example, for de Sitter JT gravity coupled to a topological gauge theory (see~\cite{Iliesiu:2019lfc} for recent work on the Euclidean AdS version of this theory). In that case the Hilbert space of asymptotic states is labeled by a number of boundaries, along with an $\ell$ and a representation of the gauge group for each boundary. Indeed, the Hilbert space decomposes into a direct sum of superselection sectors, and we expect that the annulus amplitude is given by~\eqref{E:symprop1} within each sector.

We would also like to understand the importance of the double-scaling limit. On the matrix model side we may do this by considering a generalization of the JT matrix model with a cut $[0,c]$ in which we take $c\gg 1$ and whose density of states goes over to the form~\eqref{E:JTdos} in the $c\to\infty$ limit. For example, we may consider a density of states
\beq
\label{E:backingOffJT}
	\rho_0(E) \propto \sinh\left(\sqrt{\frac{\pi E}{G}\left(1-\frac{E}{c}\right)}\right)\,.
\eeq
Now in gravity it is not at all clear if there is any sense in which we can consider large but finite $c$. But in the matrix model we may consider the basic observable $\left\langle \text{tr}\left(e^{i\ell H}\right)\text{tr}\left( e^{-i \ell' H}\right)\right\rangle_{\rm MM,\,conn}$, whose leading order behavior is a universal function of the cut and may be be reconstructed from~\eqref{E:R02}. In the $c\gg 1$ limit it is given by
\beq
\label{E:backingOff1}
	\left\langle \text{tr}\left(e^{i \ell H}\right)\text{tr}\left(e^{-i\ell'H}\right)\right\rangle_{\rm MM,\,conn} \simeq \frac{i}{2\pi}\frac{\sqrt{\ell}\sqrt{\ell'}}{\ell-\ell'+i\epsilon}  - \frac{1}{ c}\frac{1}{8\pi \sqrt{\ell}\sqrt{\ell'}}  + O(c^{-2},d^{-2})\,.
\eeq
In the $1/c$ expansion there is a singular discontinuity as one dials the imaginary part of $\ell$ from negative to positive, given by the same result $\sqrt{\ell}\sqrt{\ell'}\delta(\ell-\ell')$ we found above. However this result is misleading: for finite $d$, the left-hand-side is an integral over finite-dimensional matrices and so is finite for all $\ell,\ell'$, in which case the pole above is only approximate. To get some insight into the physics at finite $d$, we have numerically calculated the left-hand-side for a matrix model with quadratic potential. We tuned the parameters of the potential so that as $d\to\infty$ limit the model has a cut along $[0,c]$, and then numerically studied the model with large $d\gg c\gg 1 $.  For $|\ell-\ell'|\gg 1/c$ the left-hand-side of~\eqref{E:backingOff1} is well-approximated by the first term on the right-hand-side, but for $|\ell-\ell'|$ of order $1/c$ the pole is resolved into a large but finite function of height $O(c)$. In other words, at least in the quadratic ensemble, the pole and so also the singular discontinuity are resolved by finite $c$.\footnote{Correspondingly our numerical results are consistent with the existence of a pole in the double-scaled limit, non-perturbatively in $e^{-S_0}$.} 

Let us suppose that this is true for finite $c$ versions of the JT matrix model like~\eqref{E:backingOffJT}. In our dictionary, the inner product of asymptotic states in gravity came from the singular discontinuity. But if at large but finite $c$ this discontinuity is only approximate, then we see that a Hilbert space only emerges from the matrix model in the double-scaled limit. 
 
We conclude this Section with some comments on the horizon entropy of de Sitter JT gravity and the double scaled limit. As we commented in Subsection~\ref{S:entropy}, there is a semiclassical entropy of the cosmological horizon $S_{\rm cosmo} \approx 2S_0$. The finite de Sitter entropy is often cited as a hint that the Hilbert space of de Sitter quantum gravity is finite-dimensional, with a dimension given by $e^{S_{\rm cosmo}}$. Here in JT gravity we have a dual formulation in which we can attempt to assess this claim. As a measure for the dimension of the Hilbert space, let us endeavor to compute the trace of the infinite-time evolution operator $\U$, counting the number of asymptotic states that can evolve from the past to future.\footnote{We remind the reader that in JT gravity $\U$ is Hermitian on account of time-reversal symmetry. Moreover, to leading order in the genus expansion, the eigenvalues of $\U$ are either $0$ or $1$. Physically this is the statement that in the momentum basis initial states have either zero or unit probability of evolving to the infinite future.} Since we are interested in counting the number of states in a single universe, let us consider the contribution to the trace of $\U$ from 1-universe states. In the genus expansion we saw that on one-universe states of fixed momentum $\U$ acts as $\Theta(\hat{p}) + O(e^{-2S_0})$. The na\"{i}ve trace is then $\int_{-\infty}^{\infty} dp\, \bra{p}\U\ket{p}\simeq \delta(0)\int_0^{\infty}\,dp$ which has both an ``infrared'' divergence $\delta(0)$ arising from the in and out momenta being close together, and an ``ultraviolet'' divergence coming from an integral over the infinite cut. Eigenvalue repulsion leads to a rapidly oscillating phase $\sim e^{i e^{S_0}}$ on top of this leading order result, effectively smearing the in and out momenta over a scale $e^{-S_0}$. This regulates the infrared divergence, sending $\delta(0)\to e^{S_0}$. However there is still the divergent integral over the cut. If we now back off of the double-scaling limit and take the length of the cut $c$ to be large but finite, then we find an approximate but crucially finite ``trace'' goes as $\sim c \,e^{S_0}$.  If $c \sim e^{S_0}$, then $\log(c \, e^{S_0}) \approx 2 S_0$, which agrees with the semiclassical entropy.  Of course, this finding is just numerology -- the $c\sim e^{S_0}$ scaling was chosen to arrive at the pleasing answer $2 S_0$.

A key point is that by backing off the double-scaling limit, we no longer have a Hilbert space of asymptotic states, at least according to our dictionary.  Accordingly there is no ``dimension of the Hilbert space'' to speak of, although ostensibly the finite quantity $\approx \log(c \,e^{S_0})$ in the previous paragraph is some proxy for entropy.  On the other hand, in the double-scaling limit there is an exact Hilbert space of asymptotic states. We conclude that the subspace of asymptotic states which evolve to the future is infinite-dimensional.

Perhaps this should not be surprising. In JT gravity we can prepare stable de Sitter geometries, in which we can evolve for an infinite time and distinguish between final states with arbitrarily close $\ell$'s. This leads to an infinite-dimensional space of asymptotic states. However if we are limited to perform measurements over a long but finite time $\sim e^{S_0}$, then we can only reliably measure the $\ell$ of the final state to a precision $e^{-S_0}$. This effectively leads to an ultraviolet cutoff of $\sim e^{S_0}$ on the largest momenta we can resolve. The number of distinguishable asymptotic states would then be of order $e^{2S_0}$, whose logarithm $2S_0$ matches the semiclassical entropy. It would be interesting to make this precise, or in any case to establish a dictionary between the horizon entropy and a matrix model observable.

\section{Discussion}
\label{S:discussion}

In this work we have studied quantum mechanical features of de Sitter JT gravity and their avatars in its dual matrix model. JT gravity is sufficiently simple that we can compute transition amplitudes exactly as a function of the gravitational coupling $G$ as well as the sum over topologies to any desired order in the genus expansion. We consider transitions between asymptotic states, and so sum over geometries which connect past to future infinity. The computational simplicity of JT gravity has allowed us to uncover the Hilbert space of asymptotic states and evolution operator from the path integral. Our main results are:
\begin{enumerate}
	\item The asymptotic states of the model, labeled by the number of boundary components and a renormalized signed length $\ell$ for each boundary, comprise a Hilbert space. This Hilbert space is isomorphic to the Fock space of one-dimensional non-relativistic bosons. The inner product is computed by a path integral over patches of de Sitter in the infinite past or infinite future. The inner product is non-negative and non-degenerate. Unlike in ordinary quantum field theory this path integral is non-trivial, and we find a result which is not corrected in the genus expansion.
	\item This Hilbert space is equipped with three basic operators: $\hat{\ell}$ is akin to position, and measures the signed length of the boundary; its canonical conjugate $\hat{p}$, the momentum in the boundary Schwarzian modes; the commutator of $\hat{\ell}$ and $\hat{p}$ is ($-i$ times) the number operator, counting the number of universes at conformal infinity.
	\item The Hartle-Hawking wavefunction of the no-boundary state of the model is non-normalizable, and for this reason we project it out under infinite-time evolution as in Eq.~\eqref{E:projectHH}.
	\item The infinite-time path integral computes unnormalized matrix elements of the infinite-time evolution operator $\U$. To leading order in the genus expansion $\U$ is a projector, acting as the identity on positive momentum states and annihilating negative momentum states. Semiclassically, negative momentum initial states evolve into crunching cosmologies, which have vanishing probability to survive into the asymptotic future. So $\U$ acts approximately unitarily on a subspace of asymptotic states. The genus corrections to $\U$, coming from processes which change the topology of the constant time slice, are distributional terms supported at zero momentum, and these lead to further violations of unitarity.
	\item The Hilbert space of asymptotic states along with an approximately unitary $\U$ both emerge from a double-scaled matrix integral. We track both to the connected two-point function of resolvents, which is a universal function for a double-scaled matrix integral in the Gaussian Unitary Ensemble symmetry class, independent of the details of the matrix potential. Similar statements apply for the other symmetry classes. The double scaling limit is essential in our dictionary between the matrix integral averages and the inner product of asymptotic states in the gravity dual. In this sense there is an emergent Hilbert space and unitarity inside of every double-scaled matrix integral.
	\item In our dictionary the emergent Hilbert space is not coming from the eigenvalue distribution of the matrix model. Rather it comes from a certain singular discontinuity in correlation functions of resolvents. However, we do map the fact that $\U$ acts as $\Theta(\hat{p})$ on 1-boundary states, i.e.~is $1$ for $p>0$ and $0$ for $p<0$, to the statement that the dual matrix model has a semi-infinite cut stretching from zero to infinity. Further, the genus corrections to $\U$ hint at the existence of an effect which is non-perturbative in the genus expansion, namely tunneling from positive to negative momentum states, corresponding to the non-perturbatively small but nonzero likelihood of finding an eigenvalue outside the cut of a matrix model.  This is a form of instability for de Sitter space in JT gravity.
	\item With the de Sitter entropy in mind, we endeavored to count the effective dimension of the Hilbert space of asymptotic states. We did so by looking at the subspace of asymptotic states which evolve into a smooth de Sitter geometry. Using the matrix model description we found hints that the effective dimension of 1-boundary states diverges logarithmically for one-cut matrix integrals as the length of the cut is taken to infinity, and so is infinite for JT gravity. However this does not rule out a statistical mechanical interpretation for the horizon entropy, since we are counting asymptotic states rather than bulk states. 
\end{enumerate}

We stress that in this duality, unlike in ordinary examples of AdS/CFT, not only are the spatial and temporal dimensions of the gravity dual emergent, but so is the very quantum mechanical description in terms of a Hilbert space and operators acting on it. This is different from a pervasive philosophy in the AdS/CFT community, which can be stated as follows.  In AdS/CFT, string theory on AdS emerges from a quantum non-gravitational dual CFT.  Hence the slogan, ``gravity is emergent from quantum mechanics.''  In the context of considering quantum information aspects of this correspondence, the slogan becomes ``it from qubit.''

However, we see in our matrix models that a \textit{non}-quantum description of de Sitter JT gravity is more fundamental, namely the underlying random matrix ensemble.  The quantum mechanics in question arises from universal aspects of random matrices in the double scaling limit.  Perhaps the original slogan of John Wheeler~\cite{wheeler1990information}, ``it from bit'', is more appropriate here.  It is not as of yet clear if this lesson will generalize, but it provides a tantalizing hint of a description of reality which, at least in de Sitter, may be more fundamental than quantum mechanics.

There are several ways in which our approach in this manuscript is a new take on de Sitter holography.  We have a first-principles computation of the inner product on asymptotic states which allows us to compute properly normalized transition amplitudes.  These come from a sum over geometries which interpolate between past and future infinity. So there is an asymptotic past, and we allow for any number of universes in the past and in the future. As a result, the Hartle-Hawking wavefunction does not play a privileged role in the quantum mechanics of JT gravity,\footnote{However, it is worth noting that the Hartle-Hawking wavefunction encodes the leading order density of states of the dual matrix integral, and so the entire genus expansion via topological recursion. So that wavefunction indeed plays an important role, although only behind the scenes.} and in fact as a quantum state we find it to be non-normalizable.

While some of the results of this paper are specific to JT gravity and its variations, we expect many of the lessons learned to generalize more broadly.  For example, in de Sitter it is natural to sum over geometries with any number of past and future asymptotic regions. The path integral over such geometries calculates a transition amplitude between asymptotic states labeled by any number of universes. Indeed, we stress the importance of studying quantum gravity on global de Sitter space, which played a central role in our analysis. However, it is not clear if we should include similar geometries with multiple asymptotic regions for string theory in Euclidean AdS. At a more technical level, one of our main results was a first-principles computation of the inner product on asymptotic states, which involved a path integral over large gauge transformations and an integral over moduli. We expect there to be a similar although more challenging computation for de Sitter gravity in three dimensions, building upon previous results for the path integral of Euclidean AdS$_3$ gravity~\cite{Cotler:2018zff} and for dS$_3$~\cite{Cotler:2019nbi}, and perhaps even for higher-dimensional de Sitter gravity. 

Let us conclude with a list of future directions suggested by our work.

It would be particularly interesting to study the quantum mechanics of JT gravity coupled to matter fields as in~\cite{blommaert2019clocks, saad2019late}, as well as of gravity (possibly coupled to matter) in three dimensions. In all of these cases there are no propagating gravitons, but the gravitational path integral remains non-trivial on account of the integrals over large gauge transformations and moduli. It is also tempting to speculate whether the methods of our paper would help in understanding flat space holography in two and three spacetime dimensions. While there are various attempts~\cite{barnich2010aspects, Barnich:2012aw, Barnich:2012rz, Bagchi:2013toa, Andrade:2015fna, Oblak:2016eij, Barnich:2017jgw, hijano2019flat} to make sense of flat space gravity by taking a large radius of AdS gravity, the flat-space $S$-matrix is as an observable perhaps closer to transition amplitudes in de Sitter than to correlation functions of Lorentzian CFT. 

There is some early evidence~\cite{ads3trumpet} that pure Euclidean AdS$_3$ gravity is dual to an ensemble (perhaps of two-dimensional CFTs), although it is not yet clear what this ensemble is. To the extent that dS$_3$ gravity is an analytic continuation of Euclidean AdS$_3$ gravity~\cite{Cotler:2019nbi}, it would be interesting to understand whether there is an analogue of level repulsion in that ensemble, and if as in the present manuscript it leads to approximate unitarity in de Sitter. 

Besides de Sitter JT gravity the best tested example of de Sitter holography is a duality between Vasiliev gravity in an inflating patch of dS$_4$ and the singlet sector of an $Sp(N)$ vector model of anticommuting scalars~\cite{Anninos:2011ui}. Is there a duality between Vasiliev theory in global dS$_4$ and some doubled version of the $Sp(N)$ model? If such a doubled version exists, can it be used to compute transition amplitudes and scattering in dS$_4$? Can one compute the norm on asymptotic states of Vasiliev gravity from the bulk, and if so, what is the relation with the Hilbert space proposal of~\cite{Anninos:2017eib}?

Finally, we cannot help but wonder if the problem of finding global de Sitter transition amplitudes is amenable to the techniques of the conformal bootstrap program. Consider pure gravity on global dS$_3$. In that case there are two copies of Virasoro asymptotic symmetry, one for each boundary, with the diagonal combination spontaneously broken by the geometry. The approximate central charge is large and imaginary. By the analogue of the usual AdS/CFT dictionary one has independently conserved stress tensors on each boundary, as well as the constraints from asymptotic symmetry. Can the spectrum of operators in such a theory be bootstrapped in the large $c$ limit? What about the scattering of two particles in de Sitter, which can be recast as a four-point function with two insertions on the past boundary and two in the future?

There appear to be many new fruitful directions to explore in de Sitter holography, building upon recent progress in simple models of quantum gravity and CFT.  Time will tell.

\subsection*{Acknowledgments}
We would like to thank D.~Jafferis, S.~Shenker, E.~Silverstein, D.~Stanford, G.~Turiaci, and F.~Wilczek for fruitful discussions. We are especially grateful to A.~Maloney for our previous collaboration and many useful discussions. JC is supported by the Fannie and John Hertz Foundation and the Stanford Graduate Fellowship program. KJ is supported in part by the Department of Energy under grant number DE-SC 0013682. 

\bibliography{refs}
\bibliographystyle{JHEP}

\end{document}